\documentclass[12pt]{JHEP3}
\usepackage{latexsym}
\usepackage{amsfonts}
\usepackage{amsbsy}
\usepackage{epsfig}
\renewcommand{\subsection}[1]{\addtocounter{subsection}{1}
\vspace{2.5mm}\par\noindent {\it \thesubsection . #1}\par \vspace{0.5mm} }
\mathchardef\varGamma="0100 \mathchardef\varDelta="0101 \mathchardef\varTheta="0102
\mathchardef\varLambda="0103 \mathchardef\varXi="0104 \mathchardef\varPi="0105
\mathchardef\varSigma="0106 \mathchardef\varUpsilon="0107 \mathchardef\varPhi="0108
\mathchardef\varPsi="0109 \mathchardef\varOmega="010A
\def\bfone{\relax{\rm 1\kern-.35em 1}}
\DeclareFontFamily{U}{rsf}{} \DeclareFontShape{U}{rsf}{m}{n}{<5> <6> rsfs5 <7> <8> <9> rsfs7
<10-> rsfs10}{} \DeclareMathAlphabet\Scr{U}{rsf}{m}{n} \preprint{}
\title{$N=1$ reductions of $N=2$ supergravity in the presence of tensor multiplets }
\author{Riccardo D'Auria$^\S$\\ E-mail:
\email{riccardo.dauria@polito.it}}
\author{Sergio Ferrara$^\dagger$\\ E-mail:
\email{Sergio.Ferrara@cern.ch}}
\author{Mario Trigiante$^\S$\\ E-mail:
\email{mario.trigiante@polito.it}}
\author{Silvia Vaul\`a$^\star$\\ E-mail: \email{silvia.vaula@desy.de}\\
\vspace{5pt}\\
$\S$ Dipartimento di Fisica, Politecnico di Torino \\
C.so Duca degli Abruzzi, 24, I-10129 Torino.\\ and\\
Istituto Nazionale di Fisica Nucleare, Sezione di Torino,Italy\\
$\dagger$ CERN, Physics Department, CH 1211 Geneva 23, Switzerland.\\ and\\  INFN, Laboratori
Nucleari di
Frascati, Italy.\\and \\
University of California, Los Angeles, USA\\
$\star$ DESY, Theory Group\\
Notkestrasse 85, Lab. 2a D-22603 Hamburg, Germany\\ and \\
II. Institut f\"ur Theoretische Physik der Universit\"at Hamburg,\\
Luruper Chausse 179, D-22761 Hamburg, Germany.} \abstract{We consider consistent truncations
of $N=2$ supergravites in the presence of tensor multiplets (dual to hypermultiplets) as they
occur in type IIB compactifications on Calabi--Yau orientifolds. We analyze in detail the
scalar potentials encompassing these reductions when fluxes are turned on and study vacua of
the $N=1$ phases.} \preprint{CERN-TH/2005-035\\DESY 05-036}
\begin{document}
\def\theequation{\arabic{section}.\arabic{equation}}
\newcommand{\sect}[1]{\setcounter{equation}{0}\section{#1}}
\renewcommand{\theequation}{\thesection.\arabic{equation}}
\def\op#1{\mathcal{#1}}
\def\bfone{\relax{\rm 1\kern-.35em 1}}
\def\bfnull{\relax{\rm O \kern-.635em 0}}
\def\dop{{\rm d}\hskip -1pt}
\def\mez{\frac{1}{2}}
\def\sx{\left}
\def\dx{\right}
\def\na{\nabla}
\def\imez{\frac{i}{2}}
\def\a{\alpha}
\def\b{\beta}
\def\g{\gamma}
\def\d{\delta}
\def\e{\epsilon}
\def\ve{\varepsilon}
\def\vp{\varphi}
\def\tvp{\tilde{\varphi}}
\def\t{\theta}
\def\l{\lambda}
\def\m{\mu}
\def\n{\nu}
\def\r{\rho}
\def\s{\sigma}
\def\t{\tau}
\def\z{\zeta}
\def\c{\chi}
\def\p{\psi}
\def\o{\omega}
\def\G{\Gamma}
\def\D{\Delta}
\def\T{\Theta}
\def\L{\Lambda}
\def\Pg{\Pi}
\def\S{\Sigma}
\def\O{\Omega}
\def\kt{\tilde{k}}
\def\pb{\bar{\psi}}
\def\cb{\bar{\chi}}
\def\lb{\bar{\lambda}}
\def\i{\imath}
\def\Pii{\mathcal{P}}
\def\Q{\mathcal{Q}}
\def\K{\mathcal{K}}
\def\A{\mathcal{A}}
\def\N{\mathcal{N}}
\def\F{\mathcal{F}}
\def\M{\mathcal{M}}
\def\Gi{\mathcal{G}}
\def\Ci{\mathcal{C}}
\def\oL{\overline{L}}
\def\eq#1{(\ref{#1})}
\def\ol{\overline}
\newcommand{\be}{\begin{equation}}
\newcommand{\ee}{\end{equation}}
\newcommand{\ba}{\begin{eqnarray}}
\newcommand{\ea}{\end{eqnarray}}
\newcommand{\ban}{\begin{eqnarray*}}
\newcommand{\ean}{\end{eqnarray*}}
\newcommand{\nn}{\nonumber}
\newcommand{\noi}{\noindent}
\newcommand{\fgl}{\mathfrak{gl}}
\newcommand{\fu}{\mathfrak{u}}
\newcommand{\fsl}{\mathfrak{sl}}
\newcommand{\fsp}{\mathfrak{sp}}
\newcommand{\fusp}{\mathfrak{usp}}
\newcommand{\fsu}{\mathfrak{su}}
\newcommand{\fp}{\mathfrak{p}}
\newcommand{\fso}{\mathfrak{so}}
\newcommand{\fg}{\mathfrak{g}}
\newcommand{\fr}{\mathfrak{r}}
\newcommand{\fe}{\mathfrak{e}}
\newcommand{\rE}{\mathrm{E}}
\newcommand{\rSp}{\mathrm{Sp}}
\newcommand{\rSO}{\mathrm{SO}}
\newcommand{\rSL}{\mathrm{SL}}
\newcommand{\rSU}{\mathrm{SU}}
\newcommand{\rUSp}{\mathrm{USp}}
\newcommand{\rU}{\mathrm{U}}
\newcommand{\rF}{\mathrm{F}}
\newcommand{\R}{\mathbb{R}}
\newcommand{\C}{\mathbb{C}}
\newcommand{\Z}{\mathbb{Z}}
\newcommand{\Hb}{\mathbb{H}}
\def\oL{\overline{L}}
%%%%%%%%%%%%%%%%%%%%%%%%%%%%%%%%%%%%%%%%%%%%%%%%%%%%%%%%%%%%%%%%%%%%%%%%
%%%%%%%%%%%%%%%%%%%%%%%%%%%%%%%%%%%%%%%%%%%%%%%%%%%%%%%%%%%%%%%%%%%%
%%%%%%%%%%%%%%%%%%%%%% FINE MANIFOLDWALKER MACRO %%%%%%%%%%%%%%%%%%%%
%%%%%%%%%%%%%%%%%%%%%%%%%%%%%%%%%%%%%%%%%%%%%%%%%%%%%%%%%%%%%%%%%%%%%%%
%\section{$N=2\rightarrow N=1$ truncation: generalities}
\section{Introduction}

Massive deformations of extended supergravity play an important role in the description of
superstring and M--theory compactifications in the presence of fluxes \cite{gvw,drs,tv,ps,m},
either induced by p--forms or by S-S generalized dimensional reduction \cite{ss}.
\par In these compactifications to four dimensions one often
encounters non--standard supergravities in that some of the scalars have been replaced by
antisymmetric tensor fields \cite{thv,Dall'Agata:2003yr,Sommovigo:2004vj,D'Auria:2004yi,df},
which, when fluxes are turned on, may become massive vector fields \cite{df,ls,s}. The
advantage of introducing antisymmetric tensor fields is that one can introduce two kinds of
mass--deformations, one of electric and the other of magnetic type. $N=2\rightarrow N=1$
reduction of Calabi--Yau compactifications of Type IIA and Type IIB theories \cite{Andrianopoli:2001zh,adf,l},
corresponds to Calabi--Yau orientifolds \cite{gkp,Grimm:2004uq,jl,ggjl,blt,lrs,lmrs}, and one
encounters in this context such kind on antisymmetric tensor couplings to gravity as coming
from NSNS and RR 2--forms or 4--forms. In the present paper we consider, in full detail, such
reductions, for the case of different $N=2\rightarrow N=1$ truncations which correspond to
heterotic string or Calabi--Yau orientifolds with different kind of orientifolding. The paper
is organized as follows. In section 1 we describe the $N=2$ effective supergravity as coming
from Type IIB compactifications on a Calabi--Yau 3--fold \cite{bcf}--\cite{gktt}. In section 2
to 5 we discuss the different truncations which give different $N=1$ theories, in the presence
of general fluxes. In the remaining sections we discuss the nature of the vacua, the
supersymmetric configurations, and the classification of vacua in the case of cubic
prepotentials for a given set of electric and magnetic charges.

\section{$D=4$, $N=2$ supergravity from Type IIB flux compactification}

The general $N=2$ supergravity theory coupled to tensor and vector multiplets has been
discussed in references \cite{Dall'Agata:2003yr,Sommovigo:2004vj,D'Auria:2004yi}.
 We recall the field content of the effective theory which, following the
 notations and conventions of
 \cite{Andrianopoli:1996cm,D'Auria:2004yi},
  is given by:
  \begin{itemize}
\item{the gravitational multiplet $$\left(V^a_\m,\,\p_{\m A},\,\p_\m^A,\, A^0_\m\right)\,,$$
where $A=1,\,2$ is the  ${\rm SU}(2)$ R--symmetry index of the gravitinos $\psi$, lower and
upper index referring to their left or right chirality respectively, and $V^a_\mu, A^0_\m$ are
the vierbein and the graviphoton;}
 \item{$n_V$ vector multiplets$$\left(A_\m^i,\,\l^{A  i},\,\l_A^{\bar \imath},\,z^i,\,\bar z^{\bar \imath} \right)\,,$$
  where the
 chirality convention for upper and lower R--symmetry indices A of the gauginos $\lambda$ is
 reversed, and
 $z^i, \,\, i=1\dots n_V$ are the complex coordinates of the special K\"ahler manifold $\mathcal M_{SK}$;}
 \item{ a scalar--tensor multiplet $$\left(\zeta_{\a},\,\zeta^{\a},\,q^{ u},
B_{I\m\n}\right)\,,$$ where $I=1,\cdots n_T$, label the tensor fields,
$\zeta^{\a},\,\zeta_{\a}$ are the (anti)--chiral fermions ("hyperinos") $\a=1,\dots 2n_H$ ,
transforming in the fundamental of ${\rm Sp}(2n_H)$,  and $q^u$ are the coordinates of the
manifold ${\mathcal M}_T $ associated to the scalar--tensor multiplets,
 with $u=1,\dots 4n_H-n_T$.}
\end{itemize}
  If we think of this theory as coming from standard $N=2$ supergravity, $n_H$ denotes the
  number of hypermultiplets and $n_T$ the number of quaternionic coordinates which, being
  axionic, have been dualized into antisymmetric tensors. In the following we shall consider the
  particular case of a $N=2$ theory resulting from compactification of Type IIB theory on a
  Calabi--Yau 3--fold. Therefore the scalar--tensor multiplet will contain just two
  tensors $B_{1\,\m\n}, \,  B_{2\,\m\n}$, which in the ten dimensional interpretation
  come  from the ten dimensional NSNS and RR 2-forms respectively. Therefore in the following we set
   $n_T=2$ so that $I=1,\,2$. Note that in our conventions the index $I=1$ for the charges are
   related to RR fluxes while the index $I=2$ to NSNS fluxes:
   \begin{eqnarray}
e^{1}{}_{\bf \Lambda},\,m^{1{\bf \Lambda}}&\leftrightarrow & \mbox{RR fluxes}\,,\nonumber\\
e^{2}{}_{\bf \Lambda},\,m^{2{\bf \Lambda}}&\leftrightarrow & \mbox{NS fluxes}\,.
   \end{eqnarray}
  The Lagrangian and transformation laws of the theory have been given in reference
   \cite{D'Auria:2004yi}.\\ The analysis of the truncation of such theory to $N=1$ can be done
   by a careful investigation of the supersymmetry transformation laws, which are given below (up to 3--fermion terms):
\begin{eqnarray}
\d\psi_{A|\m} &=& \nabla_\m \ve_A -M^{IJ}\tilde H_{J\m} \o_{I\, A}{}^B \ve_B + \sx[  i
S_{AB}\eta_{\m\n}+\e_{AB} T^-_{\m\n}\dx]
\g^\n\ve^B\,, \label{trasfgrav0}\\
\d\l^{kA}&=& i \partial_\m z^k  \g^\m \ve^A +G^{k-}_{\m\n} \g^{\m\n} \ve_B \e^{AB}+
W^{kAB}\ve_B
\label{gaugintrasfm0}\,,\\
\d\z_\a&=& iP_{uA\a} \partial_\m q^u \g^\m \ve^A-  i M^{IJ}\tilde H_{J\m} \op{U}_{IA\a} \g^\m
\ve^A + N_\a^A
\ve_A\label{iperintrasf0}\,,\\
\d V^a_\m &=& - i \ol \psi_{A \m} \g^a \ve^A - i\ol \psi^A_\m \g^a
\ve_A\label{vieltrasf0}\,,\\
\d A^{\bf\L}_\m &=& 2 L^{\bf\L} \ol \psi^A_\m \ve^B \e_{AB} + 2 \ol L^{\bf\L} \ol \psi_{A\m}
\ve_B \e^{AB} +\nn\\
&&+\sx( i f^{\bf\L}_k \ol \l^{kA} \g_\m \ve^B \e_{AB} + i \ol f^{\bf\L}_{\bar k} \ol\l^{\bar
k}_A \g_\m \ve_B \e^{AB}\dx)
\label{gaugtrasf0}\,,\\
\d B_{I\m\n} &=& - \imez \sx(\ol \ve_A \g_{\m\n} \z_\a \op{U}_I{}^{A\a} -
\ol \ve^A \g_{\m\n} \z^\a \op{U}_{IA\alpha}\dx)+\nn\\
&& -\o_{I C}{}^{A} \sx( \ol\ve_A \g_{[\m}\psi^C_{\n]}+ \ol\psi_{[\m A}
\g_{\n]} \ve^C\dx) \label{trasfb}\,,\\
\d z^k &=& \ol \l^{kA}\ve_A \label{ztrasf0}\,,\\
\d \ol z^{\bar k}&=& \ol\l^{\bar k}_A \ve^A\,, \label{ztrasfb0}\\
\d q^u &=& P^u{}_{A\a} \sx(\ol \z^\a \ve^A + \mathbb{C}^{\a\b}\e^{AB}\ol \z_\b \ve_B
\dx)\,.\label{qtrasf0}
\end{eqnarray}
Here and in the following we give, for the sake of simplicity, the transformation laws for the
left--handed spinor fields only. \par
Notations are as follows:
\begin{itemize}
\item
 we have collected  the $n_V+1$ vectors  into
$A^{\bf\L}_\m=(A^0_\m,\,A^i_\m)$, with ${\bf\L}=0,\dots,\, n_V$, and we have
defined\footnote{We use boldface indices for the $N=2$ vector multiplets since we want to
reserve the plain capital Greek letters $\L,\,\S\dots$ to label the $N=1$ vector multiplets.}:
\begin{eqnarray}
 \tilde H_{I\m}&=&\e_{\m\n\r\s}H_I^{\n\r\s};\quad
   H_{I\,\m\n\r}=\partial_{[\m}B_{|I\,\n\r]}\,;
   \end{eqnarray}
\item
 the covariant derivative on the $\varepsilon$ parameter is given by:
\begin{eqnarray}\nabla_\m
\ve_A&\equiv&\partial_\m\ve_A-\frac14\o^{ab}_\m\g_{ab}\ve_A+\frac i2Q_\m\ve_A+\o_{\m A}^{\ \
B}\ve_B\,,\\
 Q_\m&\equiv& \frac{\rm i}{2}\left(\partial_i K\,\partial_\mu z^i\,-\,
 \partial_{\bar \imath}K \,\partial_\mu {\overline z}^{\bar \imath}\right)\,,\\
 \o_A^{\ B}&=&
 \frac {\rm i}{2}\o^x\,\s_A^{x\,B}\,,\\
\end{eqnarray}
where $\o^{ab},\,\,Q(z,\bar z),\,\,\o_A^{\,B}(q^u) $ denote the Lorentz,  {\rm U}(1)--K\"ahler
and {\rm
 SU}(2)  1--form connections, respectively.
   Here $ K$ is the special geometry
 K\"ahler potential;
 \item
 the transformation laws of the fermions (\ref{trasfgrav0}), (\ref{gaugintrasfm0}), (\ref{iperintrasf0}) also
 contain the further structures $S_{AB},\,W^{kAB},\,N_\alpha^A,\,$ named ``fermion shifts'' (or generalized
 Fayet--Iliopoulos terms) which are related to the presence
 of electric and magnetic charges $(e^I_{\bf \L}\,, m^{I\,{\bf \L}})$, and which give rise to a non trivial scalar potential.
 Their explicit form is:
\ba \label{esse1}S_{AB}& =& \frac{ i}{2}\sigma^x_{AB}\o^x_I( L^{\bf\Lambda} e^I_{\bf\L}-M_{\bf\L}m^{I\bf\L})\,,\\
\label{vudoppia} W^{kAB}&  =& i g^{k\bar \ell}\sigma^{AB}_x \o^x_I(\bar f_{\bar \ell}^{\bf\Lambda}-\bar h_{\bar\ell\bf\L}m^{I\bf\L})\,,\\
\label{enne}N_\alpha^A & =& 2\mathcal{U}^\alpha_{AI}(\bar L^{\bf\Lambda}e^I_{\bf\Lambda}-\bar
M_{\bf\L}m^{I\bf\L})\,; \ea
\item
 besides $ \tilde
H_{I\m}$ the transformation laws contain additional $I$--indexed structures ($I=1,2$), namely
 $\op{U}_I{}^{A\a}(q^u)$, $\o_{IA}{}^B(q^u)$ and $M_{IJ}(q^u)$  ($ M^{IJ}$
 will denote its inverse matrix), which satisfy a number of relations
  that can be found in ref \cite{Dall'Agata:2003yr} . We observe that,
   if one thinks of this
theory as coming from the $N=2$ standard supergravity \cite{Andrianopoli:1996cm},  the
previous $I$--indexed quantities can be interpreted as the remnants of the original  vielbein
$\op{U}^{A\a}_{\hat u}$, of the SU$(2)$ 1--form connection $\o_{\hat{u}A}^{\ \ B}$ and of the
quaternionic metric in the $I,\,J$ directions,
 after dualization of the axionic $q^I$ coordinates ($q^{\hat u} = (q^u, q^I)$) parametrizing
 the original
quaternionic manifold;
\item
 the quantity $P_{A\a}=P_{uA\a}(q^u)dq^u$ appearing in equations
\eq{iperintrasf0}, \eq{qtrasf0} is a ``rectangular vielbein'' \cite{Dall'Agata:2003yr} related
to the metric $g_{uv}$ of $\mathcal M_T$
 by the relation $P_u{}^{A\a}P_{vA\a} = g_{uv}$, and $P^{uA\a}= g^{uv}P_v{}^{A\a} $. It is related to the original vielbein
 $\mathcal{U}^{A\alpha}_{\hat{u}}$ by:
 \begin{eqnarray}
P_u{}^{A\a}&=&\mathcal{U}^{A\alpha}_{u}+A^I_u\,\mathcal{U}^{A\alpha}_I\,,
\end{eqnarray}
where $A^I_u=M^{IJ}\,h_{Ju}$ and $h_{\hat{u}\hat{v}}$ is the original quaternionic metric.
Since the quaternionic vielbein satisfies the reality condition
$\mathcal{U}^{A\alpha\star}=\epsilon_{AB}\mathbb{C}_{\alpha\beta}\,\mathcal{U}^{B\beta}$,
$\mathbb{C}$ being the ${\rm Sp}(2\,n_H)$ invariant metric, an analogous reality condition
holds for $P^{A\a}$;
 
\item
 all the other
structures appearing in the transformation laws depend on the scalar fields $z^i,\,z^{\bar
\imath}$ of the special geometry of the vector multiplets. Here we just recall the fundamental
relations obeyed by the symplectic sections of the special manifold:
\begin{eqnarray}
 D_i V &=& U_i\,,\nonumber\\
D_i U_j &=& i C_{ijk} g^{k\bar k}\bar U_{\bar k}\,,\nonumber\\
D_i U_{\bar\jmath} &=& g_{i \bar\jmath } \bar V\,,\nonumber\\
D_{i} \bar V &=& 0\,, \label{geospec}
\end{eqnarray}
where $i,\,j=1,\dots, n_V$, $D_i$ is the K\"ahler and (generally) covariant derivative, and
\begin{equation}V = (L^{\bf \Lambda},M_{\bf \Lambda})\ ,\quad
  U_i =D_i V=(f_i^{\bf \Lambda},h_{{\bf\Lambda} i})\ \ \ \ \
{\bf \Lambda}=0,\ldots,n_V \label{sezio}; \end{equation}
\begin{equation}
M_{\bf\L} = \mathcal N_{\bf\L\S} L^{\bf\S}\, , \quad h_{{\bf\Lambda} i}=\overline \mathcal
N_{\bf\L\S}f_i^{\bf \Lambda}\,,
\end{equation}
and $\mathcal N_{\bf \L\S}$ is the kinetic vector matrix. Then the ``dressed''
field--strengths $T^-_{\m\n}$ and $G^{k-}_{\m\n}$ appearing in the transformation laws of the
gravitino and gaugino fields are given by:
\begin{eqnarray}
T^-_{\mu\nu} &=& 2{\rm i} {\ Im} {\cal N}_{{\bf \Lambda}{\bf \Sigma}} L^{{\bf \Sigma}} F_{\mu\nu}^{{\bf \Lambda} -}\,,\\
G^{i-}_{\mu\nu} &=& - g^{i\bar\jmath}
 \bar f^{\bf \Gamma}_{\bar\jmath}
{\rm Im } {\cal N}_{{\bf \Gamma\Lambda}}
  {F}^{{\bf \Lambda} -}_{\mu\nu}.
\end{eqnarray}
\item Finally the scalar potential of the theory can be computed from the shifts \eq{esse1},
\eq{vudoppia},
 \eq{enne}and is given by:
\begin{eqnarray}
\label{eq:pot} \op{V}&=& 4 \left( \op{M}_{IJ} - \o^x_I \o^x_J \right) \left( m^{I{\bf \L}}
\ol{M}_{\bf \L} -e^I_{\bf \L} \ol{L}^\L \right) \left( m^{J{\bf \S}}M_{\bf \S}
-e^J_{\bf \S} L^{\bf \S} \right) + \nn\\
&&+ \o_I^x \o^x_J\left( m^{I{\bf \L}} , e^I_{{\bf \L}} \right) \op{S}
\begin{pmatrix}{m^{J{\bf \S}} \cr e^J_{{\bf \S}}}\,,\end{pmatrix}
\end{eqnarray}
\noi
 where the matrix $\op{S}$ is a symplectic matrix given explicitly by:
\begin{equation}\label{esse}
\op{S} = -\mez \begin{pmatrix}{ I_{{\bf \L\S}} + \sx( R I^{-1} R \dx)_{{\bf \L\S}} & -\sx( R
I^{-1} \dx)_{{\bf \L}}{}^{{\bf \S}}\cr -\sx(I^{-1}R\dx)^{{\bf \L}}{}_{{\bf \S}} & I^{-1|{\bf
\L\S}}}\end{pmatrix}.
\end{equation}
\noi  where $R_{{\bf \Lambda\,\Sigma}}$ and $I_{{\bf \Lambda\,\Sigma}}$ denote ${\rm Re {\cal
N}_{{\bf \Lambda}\,\bf \Sigma}}$ and ${\rm Im}{\cal N}_{{\bf \Lambda\,\bf \Sigma}}$
respectively . Furthermore the electric and magnetic charges must satisfy the the
``generalized tadpole condition'':
\begin{equation}\label{tad}
    e^I_{\bf \L} m^{J{\bf \L}} - e^J_{\bf \L} m^{I{\bf \L}} = 0\,,
\end{equation}
as a consequence of the supersymmetry Ward identity of the scalar potential and/or the
invariance of the Lagrangian under the tensor--gauge transformation:
\begin{equation}\label{tensor}
\d B_{I\,\mu\nu}= \partial_{[\mu}\Lambda_{I\nu]}; \quad \quad \d A_\mu^{\bf \Lambda} =
-2m^{{\bf \Lambda}\,I}\Lambda_{I \mu}.\end{equation} $\Lambda_{I\mu}$ being an arbitrary
vector.
\end{itemize}
In the following we shall be concerned with a theory coming from Type IIB compactification on
a Calabi--Yau 3--fold. In this case the first term is zero, due to the peculiar properties of
the special geometry derived from a cubic prepotential, so that the scalar potential contains
only the second term, namely:
\begin{equation}\label{redpot}
\mathcal V =\o_I^x \o^x_J\left( m^{I{\bf \L}} , e^I_{{\bf \L}} \right) \op{S}
\begin{pmatrix}{m^{J{\bf \S}} \cr e^J_{{\bf \S}}}\,.\end{pmatrix}
\end{equation}

\section{Conditions for a consistent $N=2\rightarrow N=1$ truncation}
We know that in a special quaternionic  manifold $\mathcal M_Q$ we
can always identify a universal hypermultiplet which is uniquely
selected by the isometries of $\mathcal M_Q$ \cite{fs}. In the
dualized theory we are considering, the universal hypermultiplet
becomes a double tensor multiplet $$\left(
B_{1\m\n},\,B_{2\m\n},\,C_0,\,\varphi\right)\,,$$ where $C_0$ is
the ten dimensional axion of Type IIB theory, $\varphi$ is the
four dimensional dilaton and $ B_{1\m\n},\,B_{2\m\n}$ are the the
four dimensional 2--forms coming from the NSNS and RR two
forms of the Type IIB
theory.\\
The possible truncations to $N=1$ can be obtained
 setting to zero a linear combination of the supersymmetry
parameters $(\ve_1,\,\ve_2)$. It is easy to see that there
 are three essentially different truncations, all the others
being equivalent, modulo ${\rm SU}(2)$
 rotations. Two of them  will be seen to correspond  to $\mathbb{Z}_2$
orientifold projections of type IIB supergravity on a Calabi--Yau 3--fold, while the third one
corresponds to the same compactification of heterotic string. 

 To understand why we have three different truncations, let us start with the simplest choice,
  following the guidelines of \cite{Andrianopoli:2001zh},
that is, let us set to zero the parameter $\ve_2$: \be\ve_2=\p_{2\m}=0\label{ep2=0}\,.\ee
Considering the surviving $\p\ve$ currents in the supersymmetry transformation laws of the
tensors \eq{trasfb}: \be\d B_{I\m\n}=\frac
i2\o^{(3)}_I(\overline\ve_1\g_{[\m}\p_{\n]}^1+\overline\ve^1\g_{[\m}\p_{\n]1})+\dots\ee we
recognize that in order to truncate one or both of the two tensors $B_I$, we have to set to
zero the corresponding structure $\o^{(3)}_I$. As we have six $\o^x_I$, with $I=1,\,2$,
$x=1,\,2,\,3$, by means of an ${\rm SU}(2)$ transformation we can always set to zero three of
them. A possible choice is the one given in reference \cite{fs}, that is in our notations:
\begin{eqnarray}\label{conn}
 \o_I^{(1)}&=&- \frac {1}{2}e^{2\varphi}\left(\matrix{0\cr  {\rm Im}\t}\right)\,\,\,;\,\,\,\,\,
  \o_I^{(2)}=\left(\matrix{0\cr
 0}\right)\,\,\,
;\,\,\,\,\,\quad \o_I^{(3)}=-\frac {1}{2} e^{2\varphi}\left(\matrix{1\cr \,{\rm
Re}\,\t}\right)\,,
\end{eqnarray}
where $\tau=-C_0+4\,i\,e^{-\varphi+\frac{K_Q}{2}}$, $K_Q$ being
the K\"ahler potential of the special K\"ahler manifold contained
in the quaternionic--K\"ahler manifold $\mathcal M_Q
$, and $\varphi-\frac{K_Q}{2}$ is the ten--dimensional dilaton.\\
 If we want to consider other possible truncations we have to set to zero  different
 combinations
of $(\ve_1,\,\ve_2)$. For this purpose we can act with a rigid ${\rm SU}(2)$ transformation on
the theory and then set to zero the new $\ve_2$ parameter. There are essentially two more
different possibilities which fulfill our requirements, which are obtained by means of a
rotation of $\theta=\frac\pi2$ on the $(x=1,x=3)$ and $(x=2,x=3)$ planes in $\mathbb{R}^3$
respectively. They correspond to setting to zero $\ve_2^\prime$ or $\ve_2^{\prime\prime}$,
namely: 
\be\ve_2^\prime =\frac{1}{\sqrt2}(-i\ve_1+\ve_2)=0\,,\label{ep2=0x}\ee 
or
\be\ve_2^{\prime \prime}=\frac{1}{\sqrt2}(-\ve_1+\ve_2)=0\,.\label{ep2=0y}\ee
It will be seen that
in these cases we obtain the $N=1$ theory corresponding to the $O(5)/O(9)$ or $O(3)/O(7)$
orientifold projection of Type IIB theory on a Calabi--Yau 3--fold, respectively.
 The corresponding values of the rotated $\omega_I^x$ are given by:
 \vskip 1cm
 $O(5)/O(9)$ case:
\begin{eqnarray}
\o_I^{(1)}=\frac {1}{2} e^{2\varphi}\left(\matrix{1\cr \,{\rm
Re}\,\t}\right)\,\,\,\,;\,\,\,\,\,  \o_I^{(2)}&=&\left(\matrix{0\cr
 0}\right)\,\,\, ;\,\,\,\,\,\quad \o_I^{(3)}=- \frac {1}{2}e^{2\varphi}\left(\matrix{0\cr
{\rm Im}\t}\right)\,.\label{conn3}
\end{eqnarray}
 \vskip 1cm
$O(3)/O(7)$ case:
\begin{eqnarray}
 \o_I^{(1)}&=&- \frac {1}{2}e^{2\varphi}\left(\matrix{0\cr
{\rm Im}\t}\right)\,\,\,;\,\,\,\,\, \o_I^{(2)}=
 -\frac {1}{2} e^{2\varphi}\left(\matrix{1\cr \,{\rm
Re}\,\t}\right)\,\,\,\, ;\,\,\,\,\,\quad \o_I^{(3)}=\left(\matrix{0\cr
 0}\right)\,,\label{conn2}
\end{eqnarray}

  Given the correspondence between the choice
of the particular supersymmetry parameter to be set to zero and of the
  values of the corresponding structures $\o^x_I$,  in order
to analyze the three cases \eq{ep2=0}, \eq{ep2=0x}, \eq{ep2=0y}, we will set $\ve_2=\p_{2\m}=0$ in equations
\eq{trasfgrav0}--\eq{qtrasf0} and then specify the connections $\o^x_I$ according
to the case \eq{conn}, \eq{conn3}, \eq{conn2} for an explicit solution of the constraints.\\
\subsection{Truncation of the gravitational multiplet}
Let us first consider the gravitino transformation law \eq{trasfgrav0} and analyze
 the consequences of the truncation
$\ve_2=0$ which do not depend of the three different choices \eq{ep2=0}, \eq{ep2=0x}, \eq{ep2=0y}.
 Following the same
steps as in \cite{Andrianopoli:2001zh}, setting $\ve_2=\p_{2\m}=0$ in equation \eq{trasfgrav0} gives,
 for $A=1$ the
supersymmetry transformation law of the $N=1$ gravitino:
\begin{equation}{\label{grav1}}
 \d\p_{1\m}=\nabla_\m\ve_1-M^{IJ}\tilde H_{J\m} \o_1^{\ 1}\ve_1+iS_{11}\g_\m\ve^1\,,
 \end{equation}
 while for $A=2$  we obtain the following
consistency condition:
\begin{equation}\label{consgrav}
\d\p_{2\m}=\o_{\m 2}^{\ \ 1} \ve_1 -M^{IJ}\tilde H_{J\m} \o_{I\, 2}{}^1 \ve_1 + \sx[ {\rm i}
S_{21}\eta_{\m\n}+\e_{21} T^-_{\m\n}\dx]\g^\n\ve^1= 0\,,
\end{equation}
 which implies:
\begin{eqnarray}
&&\o_{\m2}^{\ \ 1}= \o_{u2}^{\ \ 1}\partial_\m
q^u=0\,,\label{oconn}\\
 &&S_{21}=\frac i2\s^x_{21}\o^x_I(L^{\bf\L} e^I_{\bf\L}-M_{\bf\L}m^{I\bf\L})=0\,,\label{gauterm}\\
&&M^{IJ}\tilde H_{J\m} \o_{I\, 2}{}^1=0\,,\label{oh}\\
 &&T^-\equiv2i {\rm Im}\N_{\bf{\L\S}}L^{\bf\L} \F^{-\bf\S}=0\,.\label{gravphot}
\end{eqnarray}
The last condition can be solved as in reference \cite{Andrianopoli:2001zh} since, apart from
the fermionic shift related to the scalar potential, the vector multiplet sector is untouched
by the dualization in the hypermultiplet sector. A short account of the results given in
\cite{Andrianopoli:2001zh} is reported in the next paragraph. \\
 Conditions \eq{oh} and
\eq{gauterm} depend on the choice of one of the three aforementioned cases and will be
analyzed separately in the next sections. In this section we concentrate on those conditions
which do not depend on the choice of the structure of $\o_{IA}{}^B$.
 Condition \eq{oconn} differs from the one in reference \cite{Andrianopoli:2001zh} because here there
appears the ${\rm SU}(2)$ connection $\o^x(q^u)$ of the reduced quaternionic manifold
\cite{Dall'Agata:2003yr}, instead of the connection $\hat\o^x (q^{\hat u})$ of the
quaternionic manifold of standard $N=2$ supergravity \cite{Andrianopoli:1996cm}.\footnote {We
recall that the tensors of the scalar-- tensor multiplet come from the axionic scalars of the
$N=2$ quaternionic manifold which have been dualized implying that the residual $N=2$ manifold
is no more quaternionic} In fact,  using the expression of the  ${\rm SU}(2)$ curvature $\O_A
{}^B$ as given in \cite{Dall'Agata:2003yr}, we have:
\begin{equation}\label{consom}
\O_1 {}^2 \equiv d\o_2^{\ 1}+\o_2^{\ 1}\wedge \o_1^{\ 1}+\o_2^{\ 2}\wedge \o_2^{\ 1}+\nabla
\left(A^I\wedge\o_{I\,2}^{\ \ 1}\right)+P_{2\a}\wedge P^{1\a}\,.
\end{equation}
The consistency condition $\O_1 {}^2=0$ gives:
\begin{eqnarray}
&&\nabla \left(A^I \wedge \o_{I\,2}^{\ \ 1}\right)=0\,,\label{dof}\\
&&P_{2\a}\wedge P^{1\a}=0\,.\label{pviel}
\end{eqnarray}
%Also the condition \eq{gauterm}  differs from the corresponding one in reference
%\cite{Andrianopoli:2001zh} because of the presence of a "magnetic" contribution proportional
%to $m^{I\bf\L}$, which is absent in ordinary $N=2$ gauged supergravity, but can be introduced
%when tensors are present in the spectrum. Nevertheless we will see that the solutions of this
%constraint is a natural
 %extension of the one of \cite{Andrianopoli:2001zh}.\\
%Finally the constraint \eq{oh} is completely new since it involves the tensors and will be
%analyzed in the following when considering the three different truncations.
Since equation  \eq{dof} depends on $\o_{IA}{}^B$ it will be dealt with later. To analyze the
consequences of \eq{pviel}, we observe that the holonomy of the scalar manifold $\mathcal{
M}_T$ for the $N=2$ tensor coupled theory is contained in ${\rm SU}(2)\times{\rm
Sp}(2n_H)\otimes {\rm SO}(n_T=2)$ \cite{Dall'Agata:2003yr}. Performing the truncation from
$N=2$ to $N=1$ the holonomy must reduce according to:
\begin{equation}\label{red}
 {\rm Hol}(\M^{N=2}_Q)\subset {\rm
SU}(2)\times{\rm Sp}(2n_H)\otimes {\rm SO}(2)\rightarrow {\rm Hol}(\M^{N=1}_Q)\subset {\rm
U}(1)\times{\rm SU}(n_H)\,.
\end{equation}
We split therefore the symplectic index $\alpha$ of $P_{A\alpha}$ as follows:
\begin{equation}\label{spl}
\a\rightarrow(\hat\a,\dot\a)\in \left(\hat{\rm U}(1)\times\hat{\rm SU}(n_H)\right)\times
\left(\dot{\rm U}(1)\times\dot{\rm SU}(n_H)\right)\,.
\end{equation}
 The reality
condition on the vielbein $P^{A\a}$ becomes: \ba
P_{1\hat\a}\equiv(P^{1\hat\a})^*&=&\mathbb{C}_{\hat\a\dot\b}P^{2\dot\b}\,,\label{recon1}\\
P_{2\hat\a}\equiv(P^{2\hat\a})^*&=&-\mathbb{C}_{\dot\b\hat\a}P^{2\dot\b}\,,\label{recon2} \ea
where the symplectic metric has been decomposed according to: \be\mathbb{C}_{\a\b}=
\begin{pmatrix}{0&\mathbb{C}_{\hat\a\dot\b}\cr\mathbb{C}_{\dot\a\hat\b}&0}\,,\end{pmatrix}\ee
with $\mathbb{C}_{\hat\a\dot\b}=-\mathbb{C}_{\dot\b\hat\a}=\d_{\hat\a\dot\b}$.
 Therefore the
constraint \eq{pviel} can be rewritten as: \be\mathbb{C}_{\hat\a\dot\b}P_2^{\hat\a}\wedge
P^{1\dot\b}+\mathbb{C}_{\dot\a\hat\b}P_2^{\dot\a}\wedge P^{1\hat\b}=0\,, \ee which can be
solved setting, for instance: \be P_{2\dot\a}=0 \Leftrightarrow P_{1\hat\a}=0
\,.\label{vielzer} \ee
 Equation \eq{vielzer}, implies further constraints using the results of reference
  \cite{Dall'Agata:2003yr} for the covariant derivatives of $P_{A\alpha}$, namely:
\ba&& dP_{2\dot\a}+\o_2^{\ 1}\wedge P_{1\dot\a}+\o_2^{\ 2}\wedge P_{2\dot\a}+
\D_{\dot\a}^{\ \dot\b}\wedge P_{2\dot\b}+\D_{\dot\a}^{\ \hat\b}\wedge P_{2\hat\b}
+F^I\wedge \mathcal{U}_{I2\dot\a}=0\,,\\
&& dP_{1\hat\a}+\o_1^{\ 1}\wedge P_{1\hat\a}+\o_1^{\ 2}\wedge P_{2\hat\a}+\D_{\hat\a}^{\
\hat\b}\wedge P_{1\hat\b}+\D_{\hat\a}^{\ \dot\b}\wedge P_{1\dot\b}+F^I\wedge
\mathcal{U}_{I1\hat\a}=0\,. \ea Taking into account equations \eq{vielzer}, \eq{oconn} we
obtain the consistency constraints:
 \begin{eqnarray}
 && F^I \mathcal{U}_{I2\dot\a}=
F^I \mathcal{U}_{I1\hat\a}=0 \,,\label{fu} \\
&&\D_{\hat\a}^{\ \dot\b}= \D_{\dot\a}^{\ \hat\b}=0\,. \label{consd}
\end{eqnarray}
Furthermore, considering the curvature associated to the vanishing connections \eq{consd} it
is not difficult to see, taking into account the previous constraints, that its vanishing
implies:
\begin{equation}
\O_{\dot\a\dot\b\hat\g\dot\d}=0\,,\label{omegaquat}
\end{equation}
where $\O_{\a\b\g\d}$ is the completely symmetric tensor entering the expression of the
symplectic curvature of the quaternionic manifold $\mathcal{M}_Q $ as well in $\mathcal{M}_T $
\cite{Andrianopoli:1996cm,D'Auria:2004yi}. Note that this same constraint was obtained for the
truncation $N=2 \longrightarrow N=1$ of
the standard $N=2$ supergravity \cite{Andrianopoli:2001zh}.\\
 From the supersymmetry
transformation laws of the hypermultiplet scalars, namely: \be P_{uA\a}\d
q^u=\overline\zeta_\a\ve_A+\mathbb{C}_{\a\b}\overline\zeta^\b\ve^B\,,\ee using equation
\eq{vielzer}, we obtain that the truncated spinors of the scalar--tensor multiplet are: \be
\zeta^{\hat\a}= \zeta_{\hat\a}=0\,,\label{zetazero}\ee and imposing
$\delta\zeta^{\hat\a}=\delta\zeta_{\hat\a}=0$, namely:
\ba\delta\zeta^{\hat\a}&=&iP_u^{1\hat\a}\partial_\m q^u\g^\m\ve_1
-iM^{IJ}\tilde{H}_{\m J}\op{U}^{1\hat\a}_I\g_m\ve_1+N^{\hat\a}_1\ve^1=0\,, \label{varzeta1}\\
\delta\zeta_{\hat\a}&=&iP_{u1\hat\a}\partial_\m
q^u\g^\m\ve^1-iM^{IJ}\tilde{H}_{\m J}\op{U}_{I1\hat\a}\g_m\ve^1+N_{\hat\a}^1\ve_1=0\,, \label{varzeta2} \ea
we obtain the following further conditions: \ba
&&M^{IJ}\tilde{H}_{\m I}\op{U}_{J|2\dot\a}=M^{IJ}\tilde{H}_{\m I}\op{U}_{J|1\hat\a}=0\,,\label{zeru1}\\
&&N^{\hat\a}_1=N_{\hat\a}^1=0\,.\label{zergau} \ea Vice versa, the supersymmetry transformation
laws of the retained spinors $\zeta_{\dot\a}$ \eq{iperintrasf0} imply that the vielbein on the
reduced manifold must be related to $P_{1\dot\a}$ (and its complex conjugate $P_{2\hat\a}$),
for which the reduced torsion equation becomes:
\ba& dP_{1\dot\a}+\frac i2\o^{(3)} P_{1\dot\a}+\D_{\dot\a}^{\ \dot\b}P_{1\dot\b}+F^I\op{U}_{I|1\dot\a}=0&\,,\nn\\
&\Updownarrow&\nn\\& dP_{2\hat\a}-\frac i2 \o^{(3)} P_{2\hat\a}+\D_{\hat\a}^{\
\hat\b}P_{2\hat\b}+F^I\op{U}_{I|2\hat\a}=0\,. &\label{tors}\ea
%The last term in equation
%\eq{tors} would imply a non zero torsion for the vielbein of the reduced $N=1$ manifold. We
%will see in the following that in the case of $O(3)/O(7)$ truncation the torsion term is zero
%while in the heterotic and $O(5)/O(9)$ cases the torsion term could be different from zero.
%However we show in the Appendix that for special quaternionic manifolds where the form of
%$A^I$ and $U^I_{1 \dot \alpha}$ can be computed explicitly, also the heterotic and $O(5)/O(9)$
%cases have zero torsion thus allowing the possibility of relating the $P_{1\dot \alpha}$ to
%the vielbein of the $N=1$ manifold.
%
In the sequel we shall derive the precise relation between $P^{1\dot{\alpha}}$ and the
vielbein of the $N=1$ manifold.
\subsection{Truncation of the vector multiplets}
 As far as condition \eq{gravphot} is concerned, it can be solved exactly as in reference \cite{Andrianopoli:2001zh},
since the vector multiplet sector  is untouched by the dualization of the hypermultiplets.
Some differences arise just for the gauge terms and they are discussed in the following.
Therefore, in the sequel, we just give a short account of the derivation of the results given
in \cite{Andrianopoli:2001zh}. \\ We recall that the truncation in the vector multiplet sector
(including the graviphoton) depends on the way the constraint \eq{gravphot} is satisfied.
Denoting by $L^{\bf\L}$
 the symplectic section of the special K\"ahler manifold
$\M^{N=2}_{SK}$ of complex dimension $n_V$ of the $N=2$ standard theory, the most general
solution of the constraint is obtained by splitting the index ${\bf\L}$ of $L^{\bf\L}$ as
follows: \be \label{split}{\bf\L}=0,1,\dots,n_V\rightarrow
(X=0,1,\dots,n_C,\,\L=1,\dots,n_V^\prime)\,,\ee with $ n_C+n_V^\prime=n_V$.
 If we set $L^\L=0$ the remaining $n_C+1$ sections $L^X$ parametrize a submanifold $\M^{N=1}_V$ of
complex dimension $n_C$ of the $N=1$ scalar manifold. If we further set $\F^X_{\m\n}=0$ we satisfy the constraint
\eq{gravphot} and only $n_V^\prime$ vectors $A_\m^\L$ remain in the spectrum. According to the structure of the $N=1$
multiplets it is easy to see that $n_C$ is the number of the $N=1$ chiral multiplets and
$n_V^\prime$ the number of the $N=1$ vector multiplets.\\
Consequently we also split the index $k=1,\dots,n_V$ which labels the coordinates $(1,\,z^k)=L^{\bf\L}/L^0$ of the Special
K\"ahler manifold according to $k\rightarrow(\dot k,\,\hat k)$, where $\dot k=1,\dots,n_C$ refers to the scalars of the
chiral multiplets which parametrize the K\"ahler--Hodge manifold $\M^{N=1}_V$, while $\hat k=1,\dots,n_V^\prime$ labels
the scalars which must be truncated out.\\
According to the previous considerations the $N=2$ gaugini $\l^{kA}$ therefore decompose as
follows: \be\l^{kA}\rightarrow(\l^{\dot k 1},\,\l^{\dot k 2},\,\l^{\hat k 1},\,\l^{\hat k
2})\,. \ee Defining
\begin{equation}
\lambda_\bullet^{ \L } \equiv -2 f^{ \L }_{\hat{k}}\, \lambda^{\hat{k} 2}\,. \label{deflambda}
\end{equation}
where $ f^{ \L }_k$ is the special geometry object with a world index in the
(truncated) directions $dz^k$, it turns out that $\lambda_\bullet^{ \L }$ is the chiral gaugino
of the $N=1 $ vector multiplet such that the associated D--term is given by:
\begin{equation}\label{gau}
D^\Lambda={\rm i}\left({\rm Im}{\cal N}^{-1}\right)^{\L\S }\left(P^0_\S + P^3_\S\right)\,.
\end{equation}
%where $P^0_\S$ is the prepotentials of the Killing vectors of the Special K\"ahler manifold
%and $P^3_\S$, which for the quaternionic manifold is the third component of the prepotential
%of the Killing vectors takes in our case the form \cite{D'Auria:2004yi}:
%\begin{equation}\label{p3}
%P^3_\S= {\rm i} g^{i\bar \jmath} \s^{x\,AB} \o^x_I \sx(e^I_\L \ol{f}_{\bar\jmath}^\L - m^{I\L}
%\ol{h}_{\bar\jmath\L}\dx)
%\end{equation} $P^3_\S$.
The full analysis of reference \cite{Andrianopoli:2001zh} give furthermore a set of conditions
on the special geometry structures that are given by:
 \ba &&\F^X_{\m\n}=\Gi_{X\m\n}=0\,,\label{vec1}\\
&&L^\L=M_\L=f_{\dot k}^\L=h_{\dot k\L}=0\label{vec2}\,,\\ &&f_{\hat k}^X=h_{\hat k
X}=0\,,\label{vec2b}\\ &&\N_{X\L}=0\label{vec3}\\ && W^{\dot k 21}=W^{\hat k 11}=0\,,\label{vec4}\\
&& C_{\dot k\dot\ell\hat m}=g_{\dot k\hat{\overline{\ell}}}=0\,,\label{vec5} \ea where
$\Gi_{X\m\n}$ is the dressed field--strength dual to $\F^X_{\m\n}$, and $C_{\dot k\dot\ell\hat
m}$ and $g_{\dot k\hat{\overline{\ell}}}$ are  components of the three index tensor and of the
K\"ahler metric of the special geometry with the given particular structure of indices.

\section{The heterotic case}
Let us consider now the constraints \eq{oh},  \eq{dof}, \eq{fu}, \eq{zeru1} for the case
\eq{ep2=0} when the $\o^x_I$ are specified in equation \eq{conn}. Let us first analyze the
constraint \eq{oh}. Since $\o_{I2}{}^{ 1}=\frac i2\o^x_I\s^{x\,1}_2$, using the connections
\eq{conn}, the constraint \eq{oh} implies
\begin{eqnarray}
\tilde{H}^2_\mu&=&0\,,\label{condh}
\end{eqnarray} being
$\omega_{I=2}^{(1)}$ the only non--vanishing component of $\omega_{I2}{}^1$. Equation
\eq{condh}, explicitly reads: \be
0=\tilde{H}^2_\mu=M^{21}\tilde{H}_{1\mu}+M^{22}\tilde{H}_{2\mu}\label{ret}\,,\ee As shown in
the appendix $M^{22}\neq 0$ and $M^{12}\propto {\rm Re}\tau$, therefore the solution is given
by: \be \tilde{H}_{2\mu}=0;\quad\quad M^{12}=0 \Leftrightarrow Re\t=0\,. \label{ret1}\ee Since
\begin{eqnarray}
\nabla(A^I\,\omega_I)_2{}^1=d(A^I\,\omega_{I1}{}^2)+\omega_2{}^1\wedge
A^I\,\omega_{I1}{}^1+\omega_2{}^2\wedge A^I\,\omega_{I2}{}^1=0\,,
\end{eqnarray}
taking into account eq. (\ref{oconn}) and the fact that
  $\o_{I2}^{\ \ 1}\neq 0$ only for $I=2$, we see that
 the constraint \eq{dof} is solved if we set: \be
A^2=0\,.\label{azero}\ee Using equation \eq{condh} into equation \eq{zeru1} we obtain:
\be\op{U}_{(I=)1|2\dot\a}=\op{U}_{(I=)1|1\hat\a}=0\,.\label{uzero}\ee
Equation \eq{uzero} together with the constraint \eq{ret1} satisfies equation \eq{fu}.\\
The consistency condition:
\ba\d B_{2\m\n}&=&-\frac i2(\overline\ve_1\g_{\m\n}\zeta_\a\mathcal{U}_{(I=)2}^{1\a}-
\overline\ve^1\g_{\m\n}\zeta^\a\mathcal{U}_{(I=)2|1\a}+\nn\\
&&+\frac{i}{2}\,\o^{(3)}_2(\overline\ve_1\g_{[\m}\p_{\n]}^1+\overline\ve^1\g_{[\m}\p_{\n]1})=0\,,\label{vb}\ea
implies again: $\o^{(3)}_2\propto {\rm Re}\t=0$ and furthermore:
\be\zeta_\a\mathcal{U}_{(I=)2}^{1\a}=\zeta^\a\mathcal{U}_{(I=)2|1\a}=0\,,\label{zu}\ee which
thanks to equation \eq{zetazero} gives the following constraints:
\be\op{U}_{(I=)2|1\dot\a}=\op{U}_{(I=)2|2\hat\a}=0\,.\label{zeru2}\ee
 We now consider the conditions on the fermionic shifts. Let us start with the gravitino
shift \eq{gauterm}, where we take into account condition ${\rm Re}\tau=0$ and equation
\eq{vec2}. Then we have: \be S_{12}=\frac {i} {4} e^{2 \varphi}(L^Xe^1_X-M_Xm^{1X})=0\ee which
implies: \be e^1_X=m^{1X}=0\label{fluzero1}\ee The conditions on the hyperino shifts
\eq{zergau}, are satisfied in virtue of condition ${\rm Re}\tau=0$ and equations \eq{zeru2},
\eq{fluzero1}. Finally the condition from the gaugino shift \eq{vec4} is satisfied if we set:
\be f^\L_{\hat k}e^2_{\L}-h_{\L\hat k}m^{2\L}=0\,,\label{fluzero2}\ee which implies that
\begin{eqnarray}
e^2_{\L}&=&m^{2\L}=0\,.
\end{eqnarray}
The manifold $\M^{N=1}_T$ obtained  from the reduction of the scalar--tensor multiplet has
$2n_H-1$ dimensions and must be the product of a K\"ahler manifold parametrized by $n_H-1$
complex coordinates and a one dimensional manifold parametrized by the scalars sitting in the
linear multiplet. \par In order to identify the vielbein of the K\"ahler--Hodge manifold and
the einbein of the linear multiplet we consider the equation \be
P_{u\a[A}\op{U}_{B]I}^\a=0\,,\label{dalla}\ee which is one of the constraints defining the
scalar tensor geometry of $\mathcal{M}_T$ \cite{Dall'Agata:2003yr}. We introduce $n_H-1$
complex coordinates $w^s$ ($s=1,\dots, n_H-1$) and one real coordinate $w^0=\bar{w}^0$ and
proceed as in reference \cite{Andrianopoli:2001zh} setting: \ba
P_{u1\dot\a}dq^u&=&\frac{1}{\sqrt2} P_{s\dot\a}dw^s\,\,\,;\,\,\,\,
P_{u2\hat\a}dq^u=\frac{1}{\sqrt2} P_{\bar{s}\hat\a}d\bar{w}^{\bar s}\quad\quad\quad
(s=0,\dots,n_H-1)\label{redviel2} \ea Thanks to equations  \eq{redviel2}, the $N=2$
relation \cite{Dall'Agata:2003yr}: \be P_{uA\a}P_v^{A\a}=g_{uv}\,,\ee reduces to: \be
P_{s\dot\a}P_{\bar r}^{\dot\a}=g_{s\bar r}\,,\label{newmetr}\ee $g_{s\bar r}$ being the metric
of $\mathcal{M}_V^{N=1}$.
In virtue of equations \eq{vielzer}, \eq{redviel2}, condition (\ref{dalla}) reduces to: \be
P_{s\dot\a}\op{U}_{I2}^{\dot\a}dw^s= P_{\bar{s}\hat\a}\op{U}_{I1}^{\hat\a}d{\bar w}^{\bar
s}\,\,\,;\,\,\,\,s=0,\dots, n_H-1\,, \label{ortnew}\ee which implies
\begin{eqnarray}
P_{0\dot\a}\op{U}_{I2}^{\dot\a}&=& P_{0\hat\a}\op{U}_{I1}^{\hat\a}\,,\nonumber\\
P_{ s\dot\a}\op{U}_{I2}^{\dot\a}&=& P_{\bar
s\hat\a}\op{U}_{I1}^{\hat\a}=0\,\,\,;\,\,\,\,(s=1,\dots, n_H-1)\,.\label{elimina4}
\end{eqnarray}
Taking also $\dot{\alpha}$ running from $0$ to $n_H-1$ we can
solve the orthogonality relation (\ref{elimina4}) by setting
$\op{U}_I^{1\dot{\alpha}}=0$ except the $\dot{\alpha}=0$ component
(see the Appendix for an explicit representation of $\op{U}$), by
taking $P_{s\dot\a=0}=0$ for $s=1,\dots, n_H-1$
 and requiring $i\,P_{0\dot\a}\op{U}_{I2}^{\dot\a}$ to be real. Note that with this position equation
 (\ref{tors}) implies that the K\"ahler-Hodge manifold $\mathcal M^{N=1}_Q$ has a torsionless vielbein
 $P_{s\dot\a}$,
 $\dot \alpha = 1,\dots,n_H-1$.
 \par
According to these considerations, the hyperini $\zeta_{\dot\a}$ will be also split into one
$\zeta=\{\zeta_{\bullet},\,\zeta^{\bullet}\}$ which belongs to the linear multiplet and
$n_H-1$ $\zeta^s=\{\zeta^s_{\bullet},\,\zeta^{\bullet s}\}$ belonging to the chiral
multiplets.\\
In summary the reduced $N=1$ theory has a $\sigma$--model given by the manifold:
\begin{equation} \label{sigma}
 {\mathcal M}_V^{(N=1)} \otimes {\mathcal M}_Q^{(N=1)}\otimes
\mathbb{R}\,,
\end{equation} where ${\mathcal M}_V^{(N=1)}$ is the K\"ahler--Hodge manifold of
complex dimension $n_C$ obtained from the reduction of the vector multiplet sector,
parametrized by the coordinates $z^{\dot k}$, ${\mathcal M}_Q^{(N=1)}$ is again a
K\"ahler--Hodge manifold of complex dimension $n_H-1$ obtained by the truncation in the
scalar--tensor sector and parametrized by the coordinates $w^s$, and $\mathbb{R}$ is the one
dimensional real manifold parametrized by the scalar $\varphi$ of the residual linear
multiplet.

 We can derive the supersymmetry transformation laws for $N=1$ supergravity coupled
to one linear multiplet performing the following identifications: \ba
&&\p_{\bullet\m}=\p_{1\m};\quad\ve_\bullet=\ve_1\,,\label{gravpar}\\
&&B_{\m\n}=B_{1\m\n}\,,\label{tens}\\
&& P_{\dot\a s}\d w^s=\sqrt2 P_{u1\dot\a}\d q^u|_{\M^{KH}}\,,\label{viel}\\
&& d\varphi = - P_{1\,\dot \alpha=0}\, \\
&&\c^{\dot k}=\l^{\dot k 1};\quad\l^\L_\bullet=-2f^\L_{\hat k}\l^{\hat k2}\,,\label{gaugini}\\
&&\zeta^s=\sqrt2P^{\dot\a s}\zeta_{\dot\a};\quad \zeta=-\zeta_{\dot\a=0};\quad s=1,\dots,\,n_H-1\,,\label{hyperini}\\
&&N^s=\sqrt2P^{\dot\a s}N^1_{\dot\a};\quad N= -N^1_{\dot\a=0}\,,\label{gaugeN} \\
&&N^{\dot k}=W^{11\dot k};\quad D^\L=2if^\L_{\hat k}W^{21\hat k}\,,\label{gaugeW}\\
&&L=S_{11}\,,\label{gaugeS} \ea where for the $N=1$ theory we denote  left and right--handed
spinors with a lower and upper dot $\bullet$  respectively. We these identifications the
supersymmetry transformation laws for the $N=1$ theory are: \ba
\d V^a_\m&=&-i\pb_{\bullet\m}\g^a\ve^\bullet+h.c.\,,\\
\d\psi_{\bullet\m}& = &\nabla_\m \ve_\bullet +i\,e^{-2\,\varphi}\, \tilde{H}_{\m}\ve_\bullet
+ iL \g_\m\ve^\bullet \label{delpsi}\,,\\
\d A^\L_\m&=&\frac{i}{2} \overline{\l}^\L_\bullet\g_\m\ve^\bullet+h.c.\,,\\
\d\l^\L_\bullet&=&\F^{(-)\L}_{\m\n}\g^{\m\n}\ve_\bullet+iD^\L\e_\bullet\,,\\
\d\c^{\dot k}&=&i\partial_\m z^{\dot k}\g^\m\ve_\bullet+N^{\dot k}\ve_\bullet\,,\\
\d\zeta^s&=&i\partial_\m w^s\g^\m\ve_\bullet+N^s\ve_\bullet\,,\\
\d z^{\dot k}&=&\overline{\c}^{\dot k}\ve_\bullet\,,\\
\d w^s&=&\overline{\zeta}^s\ve_\bullet\,,\\
\delta\varphi &=& \overline{\zeta}_\bullet\,\epsilon_\bullet+h.c.\,,\\
\delta B_{\mu\nu}  &=&
\frac{1}{4}\,e^{2\,\varphi}\,\overline{\epsilon}_\bullet\,\gamma_{\mu\nu}\,\zeta_\bullet-\frac{1}{2}\,e^{2\,\varphi}\,
\overline{\epsilon}_\bullet\,\gamma_{[\mu}\psi^\bullet_{\nu]}+h.c.\,,\\
\d\zeta_\bullet&=&i\partial_m\varphi\g^\m\ve^\bullet+2 e^{-2\,\vp} \tilde{H}_{\m}\g^\m\ve^\bullet+N\ve_\bullet\,.\\
 \ea the last three transformation laws referring to the linear multiplet fields
  $$\{B_{\mu\nu},\,\zeta_\bullet,\,\zeta^\bullet,\,\varphi\}$$.
 The term $M^{IJ}\tilde H_{J\mu}$ appearing in equations \eq{trasfgrav0}, \eq{iperintrasf0}
has been reduced using the explicit form of $M^{IJ}$ and ${\mathcal U}_{IA \alpha}$ given in
the Appendix. The covariant derivative in  \eq{delpsi} is defined as follows:
\be\nabla_\m\ve_\bullet\equiv\partial_\m\ve_\bullet-\frac14\o^{ab}\g_{ab}\ve_\bullet+\frac
i2{\bf Q}_\m\ve_\bullet\label{u1conn}\ee where \be{\bf Q}_\m=Q_\m+\o^{3}_\m\,,\label{U}\ee is
the
${\rm U}(1)$ connection on the $N=1$ K\"ahler--Hodge manifold $\M^{N=1}_V\times\M^{N=1}_T$.\\
The superpotential $L$, the spin $1/2$ fermion shifts and the $D$--term turn out to be:
\begin{eqnarray}
L&=&-i e^{\vp+\frac{K}{2}}\,(L^Xe^2_{X}-M_Xm^{2X})\,,\label{L}\\
N^{\dot k}&=&ig^{\dot{k}\overline{\dot\ell}}e^{\vp+\frac{K}{2}}\,
(\bar{f}^X_{\overline{\dot\ell}}e^2_X-\bar{h}_{X\overline{\dot\ell}}m^{2X})=2g^{\dot k \overline{\dot\ell}}\nabla_{\overline{\dot\ell}}\oL\,,\label{Nk}\\
N^s&=&2\sqrt2P^{s\dot\a}\op{U}_{(I=)2\dot\a}^1(\bar L^Xe^2_{X}-\bar M_Xm^{2X})=2g^{s\overline{s}}\nabla_{\overline{s}}\oL\,,\label{Ns}\\
N&=&2\,i e^{\vp+\frac{K}{2}}\,(\bar L^Xe^2_{X}-\bar M_Xm^{2X})=2\,\frac{\partial}{\partial\varphi}\,\oL\,,\label{No}\\
D^\L&=&-\frac{1}{2} e^{2\,\vp}({\rm
Im}\N^{-1})^{\L\S}(e^1_\S-\N_{\S\Gamma}m^{1\Gamma})\,.\label{D}
\end{eqnarray}
Let us observe that the electric and magnetic charges entering the superpotential $L$ satisfy the equation \eq{tad} identically.\\
The scalar potential can be deduced from the above fermion shifts and reads: \ba
V&=&-\frac{1}{8}\,e^{4\,\varphi}\,\left[16\,e^{-2\,\varphi+K_Q}\,(e^2_X-\ol\N_{XY}m^{2Y})({\rm
Im}\N^{-1})^{YZ}(e^2_Z-
\N_{ZW}m^{2W})+\right.\nn\\
&&\quad\quad\quad+\left.(e^1_\L-\ol\N_{\L\S}m^{1\S})({\rm
Im}\N^{-1})^{\L\G}(e^1_\G-\N_{\G\D}m^{1\D})\right]\,.\label{pothet}
\ea

\section{The $O5/O9$ case}
The reduction corresponding to the case \eq{ep2=0y} is completely analogous to the heterotic
case provided we perform in all the equations the substitution $I=1\leftrightarrow I=2$ .
Thus, for example, equation \eq{oh}, when equations \eq{conn3} are considered, gives the
constraints:
\begin{eqnarray}
\tilde{H}^1_\mu&=&0\,\,\Leftrightarrow \,\,B_{1\mu\nu}=0\label{condh1} ;\quad\quad Re\t=0
\label{bizero3}\end{eqnarray} replacing the conditions \eq{condh}, \eq{ret1}. Proceeding as in
previous section we now find that in the $O5/O9$ case we obtain that the  equations
\eq{uzero}, \eq{vb}, \eq{zu},\eq{zeru2}, \eq{fluzero1}, \eq{fluzero2} are valid provided we
perform the replacement $I=1\leftrightarrow I=2$. In particular all the considerations of the
previous section after \eq{fluzero2} for the identification of the fields of the $N=1$ theory
remain the same provided we set $B_{\mu\nu}=B_{2\mu\nu}$ and replace in the fermion shifts, \eq{L}, \eq{Nk}, \eq{Ns}, \eq{No}, \eq{D}, $e^2_X \rightarrow e^1_X,\, m^2_X
\rightarrow m^1_X$, $e^1_\Lambda \rightarrow e^2_\Lambda,\, m^1_\Lambda\rightarrow m^2_\Lambda$.
In particular, the transformation laws of the $N=1$ theory are the same except for the
gravitino and the spinor $\zeta_{\bullet}$ of the linear multiplet. Indeed the term
$M^{IJ}\tilde H_{J\mu}$ appearing in equations \eq{trasfgrav0}, \eq{iperintrasf0} gives a
different contribution due to the fact that now we have $\tilde H_{1\mu}=0$ instead of $\tilde
H_{2\mu}=0 $. Using the expression of $M^{IJ}$ in the Appendix we now obtain:
\begin{eqnarray}\label{new}
\d\psi_{\bullet\m}& = &\nabla_\m \ve_\bullet -2i\,\frac{1}{\mathcal{I}^{-1|00}}
\,{e^{-\,\varphi +K_Q/2}}
\tilde{H}_{\m}\ve_\bullet \,\\
\d\zeta_\bullet&=&i\partial_m\varphi\g^\m\ve^\bullet+ \frac{1} {\mathcal{I}^{-1|00}}
\tilde{H}_{\m}\g^\m\ve^\bullet+N \ve_\bullet\,.
\end{eqnarray}
As in the Heterotic case the electric and magnetic charges entering the superpotential $L$
satisfy the equation \eq{tad} identically.\\
Finally the scalar potential for the $O5/O9$ case
is :
\begin{eqnarray}
V&=&-\frac{1}{8}\,e^{4\,\varphi}\,\left[(e^1_X-\ol\N_{XY}m^{1Y})({\rm
Im}\N^{-1})^{YZ}(e^1_Z-\N_{ZW}m^{1W})
+\right.\nonumber\\
&&\quad\quad\quad+\left.16\,e^{-2\,\varphi+K_Q}\,(e^2_\L-\ol\N_{\L\S}m^{2\S}) ({\rm
Im}\N^{-1})^{\L\G}(e^2_\G-\N_{\G\D}m^{2\D})\right]\,.\label{potO5}
\end{eqnarray}

\section{The $O3/O7$ case}
Let us consider the truncation corresponding to set to zero $\varepsilon_2$ as given in \eq{ep2=0x} and
 with the connections $\o^x_I$ given in equation \eq{conn2}.\\
We analyze first the constraint \eq{oh}, which, thanks to the expression \eq{conn2}, gives the
conditions: \be \tilde{H}_\mu^1=\tilde{H}_\mu^2=0\rightarrow
\tilde{H}_{1\mu}=\tilde{H}_{2\mu}=0\,.\label{zerobi12}\ee The same consideration holds for
equations \eq{dof}, which is solved setting: \be F^I=A^I=0,\quad I=1,\,2\label{zerof12}\ee In
virtue of equation \eq{zerof12} also the constraint \eq{fu} is satisfied and thus the
constraint \eq{vielzer} is consistent. Equations \eq{varzeta1}, \eq{varzeta2} do not give any
constraint on the  $\op{U}_I^{A\a}$  because of equation \eq{zerobi12}. All the conditions on
the $\op{U}_I^{A\a}$ come from the supersymmetry transformation law of the tensors:
\ba\d B_{I\m\n}&=&-\frac i2(\overline\ve_1\g_{\m\n}\zeta_\a\mathcal{U}_{I}^{1\a}
-\overline\ve^1\g_{\m\n}\zeta^\a\mathcal{U}_{I1\a}+\nn\\
&&+\frac
i2\o^{(3)}_I(\overline\ve_1\g_{[\m}\p_{\n]}^1+\overline\ve^1\g_{[\m}\p_{\n]1})=0\,.\label{vbI}\ea
Since in this case $\o^{(3)}_I=0$ identically, we have to impose: \be
\op{U}_{I1\dot\a}=\op{U}_{I2\hat\a}=0\label{zeru3}\ee Finally the torsion equation \eq{tors}
for the vielbeins $P_{1\dot\a}$, taking equations \eq{elimina4} and \eq{omegaquat} into account, becomes:
\be dP_{1\dot\a}+\o^{\ 1}_1 P_{1\dot\a}+\D_{\dot\a}^{\ \dot\b}P_{1\dot\b}=0\,,\label{tors1}\ee
ensuring the absence of torsion of the K\"ahler Hodge manifold.\\
As far as the fermion shifts are concerned from the gravitino shift we have the condition
\eq{gauterm}: \be S_{12}=\frac i2\o^{(3)}_I(L^X e^I_X-M_Xm^{IX})=0\ee which is satisfied since
$\o^{(3)}_I=0$. Furthermore equation \eq{zergau} is satisfied in virtue of \eq{zeru3}. Finally
the constraint \eq{vec4} imposes: \ba&&f^\L_{\hat k}e_\L-h_{\L\hat k}m^{\L}=0
\label{zerflux12}\\&& e_\L=e^1_\L+\t e^2_\L,\ m^\L=m^{1\L}+\t m^{2\L}\,.\ea which implies that
the $\Lambda$--indexed charges must be zero:
\begin{equation}\label{con1000}
 e^1_\L= e^2_\L =m^{1\L}= m^{2\L}=0
\end{equation}
 The $N=1$ theory has in this case a
$\sigma$--model given by the product of two K\"ahler--Hodge manifolds
\begin{equation}\label{sigma2}
 {\mathcal M}_V^{(N=1)} \otimes {\mathcal M}_Q^{(N=1)}
\end{equation}
of complex dimension $n_C$ and $n_H$  respectively. \\
Performing the identifications:
 \ba
&&\p_{\bullet\m}=\p_{1\m};\quad\ve_\bullet=\ve_1\,,\label{gravpar1}\\
&& P_{\dot\a s}\d w^s=\sqrt2 P_{u1\dot\a}\d q^u|_{\M^{KH}}\,,\label{viel1}\\
&&\c^{\dot k}=\l^{\dot k 1};\quad\l^\L_\bullet=-2f^\L_{\hat k}\l^{\hat k2}\,,\label{gaugini1}\\
&&\zeta^s=\sqrt2P^{\dot\a s}\zeta_{\dot\a};\quad  s=1,\dots,\,n_H-1\,,\label{hyperini1}\\
&&N^s=\sqrt2P^{\dot\a s}N^1_{\dot\a};\quad N= -N^1_{\dot\a=0}\,,\label{gaugeN1} \\
&&N^{\dot k}=W^{11\dot k};\quad D^\L=2if^\L_{\hat k}W^{21\hat k}\,,\label{gaugeW1}\\
&&L=S_{11}\,,\label{gaugeS1} \ea the supersymmetry transformation laws of the $N=1$ theory are
 given by:
\begin{eqnarray}
\d V^a_\m&=&-i\pb_{\bullet\m}\g^a\ve^\bullet+h.c.\\
\d\psi_{\bullet\m}& = &\nabla_\m \ve_\bullet +i L \g_\m\ve^\bullet\\
\d A^\L_\m&=&\frac{i}{2} \overline{\l}^\L_\bullet\g_\m\ve^\bullet+h.c.\\
\d\l^\L_\bullet&=&\F^{(-)\L}_{\m\n}\g^{\m\n}\ve_\bullet+iD^\L\e_\bullet\\
\d\c^{\dot k}&=&i\partial_\m z^{\dot k}\g^\m\ve_\bullet+N^{\dot k}\ve_\bullet\\
\d\zeta^s&=&iP_{s\dot\a}\partial_\m w^s\g^\m\ve_\bullet+N_{\dot\a}\ve_\bullet\\
\d z^{\dot k}&=&\overline{\c}^{\dot k}\ve_\bullet\\
\d w^s&=&\overline{\zeta}^s\ve_\bullet \end{eqnarray}
 where the fermion shifts are given by:
\begin{eqnarray}
L&=&-\frac {1}{4} e^{2\varphi} (L^Xe_{X}-M_Xm^{X})\label{L12}\\
N^{\dot k}&=&- \frac {1}{2}g^{\dot{k}\overline{\dot\ell}} e^{2 \varphi}
(f^X_{\overline{\dot\ell}}\bar e_X-h_{X\overline{\dot\ell}}\bar m^{X})=
2g^{\dot k \overline{\dot\ell}}\nabla_{\overline{\dot\ell}}\oL\label{Nk12}\\
N^s&=&2\sqrt2 P^{s\dot\a}\op{U}_{I1\hat\a}(\bar L^Xe^I_{X}-\bar M_Xm^{IX})
=2g^{s\overline{s}}\nabla_{\overline{s}}\oL\label{Ns12}\\
D^\L&=&0\,,\label{D12}
 \end{eqnarray}
  and we have defined:
 \begin{equation}
 e_X=e^1_X+\t e^2_X;\quad\quad
 m^X=m^{1X}+\t m^{2X}\,.
\end{equation}
In the present case both the NSNS and RR electric and magnetic charges enter the definition
of the superpotential \eq{L12}. We see that the charges $\{e^1_X,\, e^2_X\}$ and
$\{m^{1\,X},\, m^{2\,X}\}$ are constrained by the tadpole condition:
\begin{eqnarray}
m^{1X}\e^2_X-m^{2X}\e^1_X&=&m^{1{\bf \Lambda}}\,e^2_{{\bf \Lambda}}-m^{2{\bf
\Lambda}}\,e^1_{{\bf \Lambda}}=0\,.
\end{eqnarray}
 For $e\times m=0$ the scalar potential is
given by \cite{tv}: \be V=-\frac{1}{8}\, e^{4\,\varphi} (e_X-\ol\N_{XY}m^Y)({\rm
Im}\N^{-1})^{XZ}(\ol{e}_Z-\N_{ZW}\ol{m}^W)\,.\label{potO3}\ee On the other hand, if $e\times
m\neq 0$, $N=2$ supersymmetry is broken but still the theory can have an unbroken $N=1$ sector.
Indeed for $e\times m\neq 0$ the scalar potential is given by \cite{blt}:
 \be V=-\frac{1}{8}\, e^{4\,\varphi}
(e_X-\ol\N_{XY}m^Y)({\rm
Im}\N^{-1})^{XZ}(\ol{e}_Z-\N_{ZW}\ol{m}^W)+\frac{1}{4}\,e^{4\varphi}\,{\rm Im }\tau\,m\times
e\,.\label{npotO3}\ee The potential \eq{npotO3} can be written in a manifestly $N=1$ fashion:
\begin{eqnarray}
V&=&e^{K(z,\,\bar{z})+K(\tau,\,\bar{\tau})+K_D(w,\,\bar{w})}\,\left[G^{i\bar{j}}\,D_i
W\,D_{\bar{j}} W+ G^{\tau\bar{\tau}}\,D_\tau W\,D_{\bar{\tau}} W\right]+e^{K_D(w,\,\bar{w})}\,
m\times e\,,\nonumber\\&&\label{potvt0}
\end{eqnarray}
where the superpotential $W$ has the form:
\begin{eqnarray}
W&=&X^X\,e_X-F_X\,m^X\,,\label{gvww}
\end{eqnarray}
which is consistent with the general expression given in \cite{gvw}. Note that since the first
term in \eq{potvt0} is separately $N=1$ supersymmetric, the last term should be supersymmetric
as well. In fact it is a F.I. term. A similar term arise from a ${\rm U}(1)$ gauge field on a
$D7$ brane world volume with magnetic fluxes \cite{dkvp}. This term explicitly breaks $N=2$
supersymmetry. Indeed for $m\times e \neq 0$, the $N=2$ Ward identity for the scalar potential
acquires an additional contribution from the square of the gaugino shifts, which is not
proportional to $\delta_A^B$ and has the form:
\begin{eqnarray}
\epsilon^{xyz}\,\omega_I^x\,\omega_J^y\,m^{I{\bf \Lambda}}\,e^J_{{\bf
\Lambda}}\,\sigma_A^{z\ B}&=&\frac{1}{4}\,e^{4\varphi}\,{\rm Im }\tau\,(m\times
e)\,\sigma^{3\ B}_A\,.
\end{eqnarray}
From a microscopic point of view, the potential in the form \eq{potvt0} does not take into
account the contributions due to $O3/O7$ planes. These, as discussed in \cite{blt}, have the
effect of canceling the last term, according to generalized tadpole cancellation condition.
The resulting potential will have the form in \eq{potO3} with $m\times e \neq 0$ and will in
general have non trivial vacua, as discussed in section 9.

\section{Comparison with the orientifold projection}
 Let us now
recover from the previous analysis the results of \cite{Grimm:2004uq}. For this purpose let us
write down the relations between our notations given in the Appendix and those of reference
\cite{Grimm:2004uq}.
 \ba
&&\tilde \xi_a\rightarrow\r_a\,\,;\,\,\,\tilde\xi_0 \rightarrow q^2\,,\\&& \xi^a\rightarrow
\ell\, b^a-c^a\,\,;\,\,\,
\xi^0= C_0\rightarrow \ell\,,\\
&&{\rm Re}(w^a)\rightarrow b^a\,\,;\,\,\, {\rm Im}(w^a)\rightarrow v^a\,\,;\,\,\,
a=1,\dots,h^{(1,1)}\,, \ea where $c^a$ and $b^a$ are the scalars coming from the RR and NSNS
two-form respectively, $v^a$ are the scalars coming from the deformations of the K\"ahler
class of the metric, while $\r^a$
 are the scalars coming from the RR four-form. The scalars $(q^1,\,q^2)$ in this context appear
 dualized into rank two tensors $(B^1_{\m\n},\,B^2_{\m\n})$ \cite{Dall'Agata:2003yr}  as they
 come from the NSNS and RR 2-- form respectively.\\
According to the $\mathbb{Z}_2$ orientifold projection, the quaternionic scalars may appear as the
coefficients of the expansion in $H^{(1,1)}_+$ or $H^{(1,1)}_-$ forms. Moreover the real part
of the complex dilaton $C_0$ and the NSNS and RR two forms $B_1$, $B_2$ may be even or not
under the  $\mathbb{Z}_2$ projection. With the previous considerations, we can see, analyzing the
truncation of the scalar--tensor multiplet, that the second truncation corresponds to the
$O5/O9$ planes case, since $C_0=B_1=0$, while the last corresponds to the $O3/O7$ planes
case, since $B_1=B_2=0$. Further analyzing the condition \eq{oconn} using the explicit
parametrization of \cite{fs} one can also check the consistency of the
truncation for the remaining scalars in the hypermultiplet sector.\\
The first truncation we considered  corresponds instead to the  orientifold projection of the
Heterotic string on a Calabi--Yau 3--fold. Nevertheless from the condition \eq{oconn} we can
identify which are the two sets of scalars whose indices must be orthogonal. Furthermore,
considering that if the NSNS two forms survive
then the scalars $b^a$ may be thought as the coefficients of the $ H^{(1,1)}_+$ expansion.\\
The results are summarized in the following table: \be\begin{matrix}{O5/O9&O3/O7&
\mbox{heterotic}\cr ~&~&~ \cr b^{\dot{a}},\ \r^{\dot a}\in H^{(1,1)}_-  & b^{\dot a},\ c^{\dot
a}\in H^{(1,1)}_-& c^{\dot a},\  \r^{\dot a}\in H^{(1,1)}_-  \cr c^{\hat a},\ v^{\hat a}\in
H^{(1,1)}_+ & v^{\hat a},\ \r^{\hat a}\in H^{(1,1)}_+& b^{\hat a},\ v^{\hat a}\in
H^{(1,1)}_+\cr C_0=0,\ B_1=0& B_1=0,\ B_2=0 & C_0=0,\ B_2=0}
\end{matrix}\label{summary}\ee
where $\dot a=1,\dots, h^{(1,1)}_-$, $\hat a=1,\dots, h^{(1,1)}_+$.\\
As far as the vector multiplets are concerned, before the truncation we had $h^{(2,1)}+1$ vector multiplets labeled by ${\bf\L}=0,\,1,\dots,h^{(2,1)}$. We split ${\bf\L}\rightarrow(\L,\,X)$ and retained the vectors $F^{\L}_{\m\n}$ and the symplectic sections $(L^X,\,M_X)$. It is now clear that for the $O5/O9$ case $\L=1,\dots,h^{(2,1)}_-$, in order to have $ h^{(2,1)}_-$ vector multiplets \cite{Grimm:2004uq}, while $X=0,1,\dots, h^{(2,1)}_+$ such that $L^X/L^0$ describe the scalars of $h^{(2,1)}_+$ chiral multiplets, while for the $O3/O7$ planes case $\L=1,\dots,h^{(2,1)}_+$, labels the $ h^{(2,1)}_+$ vector multiplets, while $X=0,1,\dots, h^{(2,1)}_-$ such that $L^X/L^0$ are the scalars of $h^{(2,1)}_-$ chiral multiplets.\\
Let us now consider the terms coming from the flux $G=H_2+\t H_1$.\\
For the $O5/O9$ case we have that \be H_2\in H_+^{(3)};\quad\quad H_1\in H_-^{(3)}\ee
therefore consistently we have the following fluxes: \be (e^1_X,\, m^{1X}),\quad X=0,1,\dots,
h^{(2,1)}_+;\quad\quad (e^2_\L,\, m^{2\L}),\quad \L=1,\dots, h^{(2,1)}_- \ee For the $O3/O7$
case we have that \be H_1\in H_-^{(3)};\quad\quad H_2\in H_-^{(3)}\ee therefore consistently
we have the following fluxes: \be (e^1_X,\, m^{1X}),\quad X=0,1,\dots, h^{(2,1)}_-;\quad\quad
(e^2_X,\, m^{2X}),\quad X=0,1,\dots, h^{(2,1)}_- \ee As far as the heterotic truncation is
concerned, we can rephrase equation \eq{condh} in the Calabi--Yau language, as the condition:
\be H_{1}\in H_+^{(3)};\quad\quad H_{2}\in H_-^{(3)}\ee therefore
consistently with \eq{fluzero1}, \eq{fluzero2} one obtains the following fluxes: \be
(e^1_\L,\, m^{1\L}),\quad \L=1,\dots, h^{(2,1)}_-;\quad\quad (e^2_X,\, m^{2X}),\quad
X=0,1,\dots, h^{(2,1)}_+ \ee which also means that we have $ h^{(2,1)}_-$ vector multiplets
and $ h^{(2,1)}_+$ hypermultiplets. Let us finally observe that equations \eq{potO3},
\eq{potO5} coincide respectively with the scalar potentials obtained in reference
\cite{Grimm:2004uq} for the $O3/O7$ and $O5/O9$ planes truncations,
 and that equation \eq{L12} gives the superpotential of reference \cite{gvw}.

\section{Supersymmetric configurations}
In the $N=2$ theory we the following fluxes are present: \ba
G^{(0,3)}&=&e^{-\frac {K}{2}}L^{\bf\L}(e_{\bf\L}-\N_{\bf\L\S}m^{\bf\S})\,,\label{flux03}\\
G^{(1,2)}_i&=&e^{-\frac{K}{2} }f_i^{\bf\L}(e_{\bf\L}-\overline{\N}_{\bf\L\S}m^{\bf\S})\,,\label{flux12}\\
G^{(3,0)}&=&e^{-\frac{K}{2}}\overline{L}^{\bf\L}(e_{\bf\L}-\overline{\N}_{\bf\L\S}m^{\bf\S})\,,\label{flux30}\\
G^{(2,1)}_{\overline k}&=&e^{-\frac {K}{2}}\bar{f}_{\overline
k}^{\bf\L}(e_{\bf\L}-\N_{\bf\L\S}m^{\bf\S})\,,\label{flux21}\ea where we recall that $K\equiv
K(z,\bar{z})$ is the K\"ahler potential of the complex structure moduli and \be e_{\bf\L}=
e_{\bf\L}^1+\t e_{\bf\L};\quad m^{\bf\L}=m^{1\bf\L}+\t m^{2\bf\L}\,, \ee and the flux
parameters satisfy the tadpole cancellation condition: \be e_{\bf\L}^1m^{2\bf\L}-
e_{\bf\L}^2m^{1\bf\L}=0\,.\label{tadpole}\ee The $N=1$ scalar potential can be always written
in the following form: \ba V&=&-\frac18 e^{4\vp} (e_{\bf\L}-\ol\N_{\bf\L\S}m^{\bf\S})({\rm
Im}\N^{-1})^{\bf\L\G}(\ol{e}_{\bf\G}-
\N_{\bf\G\D}\ol{m}^{\bf\D})=\nn\\
&=& \frac14 e^{4\vp+K}\left(G^{(3,0)}\overline{G^{(3,0)}}+g^{i\overline
k}G^{(1,2)}_i\overline{G^{(1,2)}}_{\overline k}\right)\,,\label{potT&V}\ea even in the case in
which $e\times m\neq 0$, as discussed at the end of section 6. The Minkowski minimum
corresponds to: \be G^{(3,0)}=G_i^{(1,2)}=0\label{minN=2}\ee
nevertheless the solutions of \eq{minN=2} are not consistent with the constraint \eq{tadpole}.\\
If we perform an $N=2\rightarrow N=1$ truncation the fluxes, according to equations \eq{vec2},
\eq{vec2b}, \eq{vec3} reduce to: \ba
G^{(0,3)}&=&e^{-\frac {K}{2}}L^X(e_X-\N_{XY}m^{Y})\,,\label{flux03r}\\
G^{(1,2)}_{\hat k}&=&e^{-\frac{K}{2}}f_{\hat k}^{\L}(e_{\L}-\overline{\N}_{\L\S}m^{\S})\,,\label{flux12ra}\\
G^{(1,2)}_{\dot k}&=&e^{-\frac {K}{2}}f_{\dot k}^{X}(e_X-\overline{\N}_{XY}m^Y)\,,\label{flux12rb}\\
G^{(3,0)}&=&e^{-\frac {K}{2}}\overline{L}^X(e_X-\overline{\N}_{XY}m^{Y})\,,\label{flux30r}\\
G^{(2,1)}_{\hat{\overline k}}&=&e^{-\frac {K}{2}}\bar{f}_{\hat{\overline k}}^{\L}(e_{\L}-\N_{\L\S}m^{\S})\,,\label{flux21ra}\\
G^{(2,1)}_{\dot{\overline k}}&=&e^{-\frac {K}{2}}\bar{f}_{\dot{\overline
k}}^X(e_X-\N_{XY}m^Y)\,.\label{flux21rb} \ea Consider now the case \eq{ep2=0x} corresponding
to the $O3/O7$ truncation. We report here the scalar potential and the superpotential are
given by equation \eq{potO3}, \eq{L12}.
\ba V&=&-\frac18 e^{4\vp} (e_X-\ol\N_{XY}m^Y)({\rm Im}\N^{-1})^{XZ}(\ol{e}_Z-\N_{ZW}\ol{m}^W)=\nn\\
&=&\frac14 e^{4\vp+K}\left(G^{(3,0)}\overline{G^{(3,0)}}+g^{\dot \ell\overline{\dot
k}}G^{(1,2)}_{\dot\ell}\overline{G^{(1,2)}}_{\overline{\dot k}}\right)\,,\label{potO3rep}\ea
\be L=-\frac 14 e^{2\vp} (L^Xe_{X}-M_Xm^{X})\,.\label{superpot}\ee We recall also the
condition \eq{zerflux12} for a consistent truncation: \be f^\L_{\hat k}e_\L-h_{\L\hat
k}m^{\L}= f^\L_{\hat k}(e_\L-\overline{\N}_{\L\S}m^\S)=0\,.\label{zerflux12r}\ee Therefore one
can observe that the condition for a Minkowski vacuum requires: \be
G^{(3,0)}=G^{(1,2)}_{\dot\ell}=0\,,\label{null3012}\ee while the vacuum is supersymmetric if
also the gravitino shift vanishes: \be L=0\leftrightarrow G^{(0,3)}=0\,. \label{null03}\ee The
condition \eq{zerflux12r} for a consistent truncation requires: \be
G^{(1,2)}_{\hat\ell}=0\,.\label{null12}\ee Therefore  the theory admits a supersymmetric $N=1$
Minkowski vacuum, just for $(2,1)$ fluxes, according to the previous analysis
\cite{Grana:2001xn,gkp}. Note that the minimum condition of the $N=1$ theory, can not impose
any constraint on the component of the $(1,2)$ flux along the truncated scalars $\hat\ell$
\eq{flux12ra}. The absence of such a component comes
from the constraint \eq{null12} for a consistent truncation.\\
 In the next section, we will show that
conditions \eq{null3012}, \eq{null03}, \eq{null12} do not admit a non trivial solution in the
charges if \eq{tadpole} holds. The only case in which $e\times m$ can be different from zero
after truncation to $N=1$ is the $O3/O7$ case, in which condition \eq{tadpole} is indeed
relaxed according to the discussion at the end of section 6. In virtue of this in the $O3/O7$
truncation conditions \eq{null3012}, \eq{null03}, \eq{null12} do admit a non--trivial
solution.
\section{Vacua of IIB on $CY_3$ orientifolds}
Let us now study the vacua of Type IIB string theory compactified
on a $CY_3$ orientifold. We start by recalling some general
properties of the scalar potential which hold in all the three
truncations considered earlier. To this end we shall write the
potential in a general $N=1$ form which will yield the expressions
in eqs. \eq{pothet}, \eq{potO5}, \eq{potO3} upon performing the
corresponding truncation on the fields and charges. The complex
3--form flux across a 3--cycle of the $CY_3$ can be expanded in a
basis of the corresponding cohomology group:
\begin{eqnarray}
G_{(3)}&=& H_2+ \tau \, H_1=
e_{\bf \Lambda}\,\beta^{\bf \Lambda}+m^{\bf \Sigma}\,\alpha_{\bf \Sigma}\,,\nonumber\\
e_{\bf \Lambda} &=& e^1_{\bf \Lambda}+ \tau\, e^2_{\bf
\Lambda}\,\,;\,\,\,m^{\bf \Lambda} = m^{1{\bf \Lambda}}+ \tau\,
m^{2{\bf \Lambda}}\,.
\end{eqnarray}
Let
\begin{eqnarray}
\Omega (z)&=& \left(\matrix{X^{\bf \Lambda}(z)\cr F_{\bf
\Sigma}(z)}\right)\,,
\end{eqnarray}
be  the holomorphic section on the Special K\"ahler manifold
depending only on the complex structure moduli $z^i$ ($i=1,\dots ,
h^{(2,1)}$). According to our previous analysis, the above
quantities can be specialized to the three truncations as follows:
\begin{itemize}
\item {\bf Heterotic case:} Set ${\rm Re}(\tau)=0$, $X^\Lambda=F_\Lambda=0$, $(\Lambda=1,\dots, h^{2,1}_-)$,
$e^1_X=m^{1X}=e^2_\Lambda=m^{2,\Lambda}=0$, $(X=0,\dots,
h^{2,1}_+)$.
\item {\bf O5/O9 case:}  Set ${\rm Re}(\tau)=0$, $X^\Lambda=F_\Lambda=0$, $(\Lambda=1,\dots, h^{2,1}_-)$,
$e^2_X=m^{2X}=e^1_\Lambda=m^{1,\Lambda}=0$, $(X=0,\dots,
h^{2,1}_+)$.
\item {\bf O3/O7 case:}  Set  $X^\Lambda=F_\Lambda=0$,
$e_\Lambda=m^{\Lambda}=0$, $(\Lambda=1,\dots, h^{2,1}_+)$.
\end{itemize}
 The general form of the  GVW superpotential in the low--energy
${\mathcal N}=1$ theory \cite{gvw} is
\begin{eqnarray}
W(\tau, z^i)=e_{\bf \Lambda}\, X^{\bf \Lambda}-m^{\bf \Sigma} \,
F_{\bf \Sigma}\,,
\end{eqnarray}
and the potential has the form:
\begin{eqnarray}
V&=&e^{K(z,\,\bar{z})+K(\tau,\,\bar{\tau})+K_D(w,\,\bar{w})}\,\left[G^{i\bar{j}}\,D_i
W\,D_{\bar{j}} W+ G^{\tau\bar{\tau}}\,D_\tau W\,D_{\bar{\tau}}
W\right]\,,\label{potvt}
\end{eqnarray}
where $K(z,\bar{z}),\,K(\tau,\bar{\tau}),\,K_D(w,\bar{w})$ are the
contributions to the K\"ahler potential of the $N=1$ manifold
related to the submanifolds parametrized by the complex structure
moduli $z^i$, the ten dimensional axion/dilaton $\tau$ and the
K\"ahler moduli in the ten dimensional Einstein frame $w^a$. By
comparing the expression of the potential \eq{potvt} with the
results obtained in the previous sections we find the following
identification:
\begin{eqnarray}
e^{K(\tau,\,\bar{\tau})+K_D}&=&\frac{1}{4}\,e^{4\,\varphi}=\frac{1}{4}\,e^{4\,\phi+2\,K_Q}=\frac{1}{4}\,e^{\phi+K_D}\,,\nonumber\\
K(\tau,\bar{\tau})&=&-\ln[-i\,(\tau-\bar{\tau})]+\mbox{const.}\,\,\,;\,\,\,K_D(w,\,\bar{w})=-2\,\ln
\left(\frac{1}{3!}\,d_{abc}v^av^bv^c\right)\,,\label{kmess}
\end{eqnarray}
where $v^a={\rm Im}(w^a)$ are K\"ahler moduli in the Einstein
frame. To understand the identifications in \eq{kmess}, recall
that from Type II string theory point of view $K_Q$ is related to
the volume of the $CY_3$ expressed in the ten dimensional string
frame:
\begin{eqnarray}
K_Q&=&-\ln \left(\frac{1}{3!}\,d_{abc}v_s^av_s^bv_s^c\right)\,,
\end{eqnarray}
where $v_s^a$ are K\"ahler moduli in the string frame. Since the
K\"ahler moduli $v$ in the two frames are related in the following
way $v_s^a=v^a\,e^{\frac{\phi}{2}}$,   if we define:
\begin{eqnarray}
K_D(w,\,\bar{w})&=&-2\,\ln
\left(\frac{1}{3!}\,d_{abc}v^av^bv^c\right)\,,
\end{eqnarray}
we have
\begin{eqnarray}
2\,K_Q&=&K_D-3\,\phi\,.
\end{eqnarray}
  The
potential V is extremized if $D_\tau W=D_i W=0$. However if in
addition we also require supersymmetry, we have to impose that
$0=D_{w^a} W$. This implies $W=0$, since $W$ is $w^a$--independent
and thus $D_{w^a} W\propto W$.\par We further note that, being
$K(\tau,\bar{\tau})=-\ln (-i(\tau-\bar{\tau}))+\mbox{const.}$:
\begin{eqnarray}
D_\tau
W&=&\frac{1}{(\bar{\tau}-\tau)}\,\overline{\hat{W}}\,\,\,;\,\,\,\hat{W}=e_\Lambda\,
\overline{X}^\Lambda-m^\Sigma \, \overline{F}_\Sigma\,,
\end{eqnarray}
the minimum conditions can then be written in the following form:
\begin{eqnarray}
D_\tau W=D_i W=0&\Leftrightarrow &
e_\Lambda-\overline{\mathcal{N}}_{\Lambda\Sigma}
m^\Sigma=0\,.\label{minimum}
\end{eqnarray}
The above equation clearly has solutions only if $m\times e > 0$.
This is a consequence of the following relation which holds at the
minimum:
\begin{eqnarray}
m\times e&=&
m^{1\Lambda}e^2_\Lambda-m^{2\Lambda}e^1_\Lambda=\frac{1}{{\rm
Im}\tau}\,{\rm Im}(\bar{m}e)= -\frac{1}{{\rm
Im}(\tau)}\,\bar{m}^T\,{\rm Im}\mathcal{N}\,m>0\,.
\end{eqnarray}
Note that in both the heterotic and the O5/O9 truncations $m\times
e=0$ and thus the potential has no non--trivial vacua. Only in the
O3/O7 case we can have $m\times e>0$. In what follows we shall
focus on this latter case and, with an abuse of notation, we shall
use the index $\Lambda$ to label the surviving charges:
$\Lambda=0,\dots, h^{2,1}_-$.\par
 Supersymmetry further requires $W=0$, namely:
\begin{eqnarray}
(e_\Lambda-\mathcal{N}_{\Lambda\Sigma}
m^\Sigma)\,X^\Lambda&=&0\,.\label{susyminima}
\end{eqnarray}
According to Michelson's analysis \cite{m} the vector of
electric/magnetic charges can always be reduced to a form defined
by the following non--vanishing entries (Michelson's basis):
\begin{eqnarray}
e_0&=&e_0^1+\tau\,e_0^2\,;\,\,\,e_1=e_1^1\,;\,\,\,m^0=m^{1\,0}\,\nonumber\\
m\times e&=&m^{1\,0}e_0^2>0\,,
\end{eqnarray}
which we shall refer to as Michelson's charge basis $Q_M$. From eqs. (\ref{minimum}) the
minimum conditions read:
\begin{eqnarray}
\overline{\mathcal{N}}_{0,0}&=&\frac{e_0^2 \tau+
e_0^1}{m^{1\,0}}\,\,;\,\,\,\,\overline{\mathcal{N}}_{0,1}=\mathcal{N}_{0,1}=\frac{e_1^1}{m^{1\,0}}\,\,;\,\,\,\,\,
\overline{\mathcal{N}}_{0,k}=0\,\,\,k\neq 0,\,1\,.\label{ncond}
\end{eqnarray}
If we also look for supersymmetric vacua should require eq. (\ref{susyminima}), namely
\begin{eqnarray}
(e_0-\mathcal{N}_{0,0} m^0)\, X^0+(e_1-\mathcal{N}_{1,0} m^0)\,
X^1&=&0\,\,\Rightarrow\,\, X^0=0\,,
\end{eqnarray}
where we have used the reality of $\mathcal{N}_{0,1}$ and the fact that
$(e_0-\mathcal{N}_{0,0} m^0)\neq 0$ (since $(e_0-\overline{\mathcal{N}}_{0,0} m^0)= 0$ and
being the imaginary part of $\mathcal{N}_{0,0}$ always non--vanishing). This minimum, having
$X^0=0$ cannot be described in the ordinary special coordinate patch in which $X^0\neq 0$.\par
In \cite{blt} supersymmetric vacua of an STU model (corresponding in the $O3/O7$ case to
$h^{2,1}_-=3$) have been studied in the special coordinate frame, making for the
electric/magnetic charges, which are eight complex in general, the following choice:
\begin{eqnarray}
Q_L&=&\{m^\Lambda,\, e_\Sigma\}=\{-1,\, 0,\, 0,\, \tau,\,-\tau,\,
0,\, 0,\, -1 \},\,\;\,\,\,\,\, m\times e=2\,.
\end{eqnarray}
The scalar fields of this model, denoted by $s,t,u$, in the
special coordinate basis are given by:
\begin{eqnarray}
s&=&\frac{X^1}{X^0}\,\,;\,\,\,t=\frac{X^2}{X^0}\,\,;\,\,\,u=\frac{X^3}{X^0}\,\,;\,\,\,X^0\neq
0\,.
\end{eqnarray}
 We can also define a
Michelson's basis $Q_M$ for the STU model in which the non
vanishing electric and magnetic charges correspond to
$\Lambda=0,\,3$. The symplectic bases $Q_L$ and $Q_M$ (in which
$m^{10}=2/e^2_0$ if we require $m\times e=2$) are related by a
symplectic matrix ${\Scr A}$ given in eq. \eq{A} in Appendix B:
\begin{eqnarray}
Q_M&=&{\Scr A}\,Q_L=\{\frac{2}{e_0^2},\, 0,\, 0,\,0,\,
e_0^1+\tau\,e_0^2,\, 0,\, 0,\, e_1^1 \}\,.
\end{eqnarray}
In Appendix B, eq. \eq{nstu}, the reader may also find the
explicit form of the period matrix $\mathcal{N}$ for the STU model
in the special coordinate frame. In the special coordinate basis,
with the choice of charges $Q_L$, conditions (\ref{minimum}) and
(\ref{susyminima}) have the following solution \cite{blt}:
\begin{eqnarray}
\tau &=&-u \,\,;\,\,\,\, s=-\frac{1}{t}\,.\label{vac}
\end{eqnarray}
Upon application of ${\Scr A}$ to $\Omega$ we obtain the holomorphic section $\Omega^\prime$
in Michelson's basis as function of $s,t,u$:
\begin{eqnarray}
\Omega^\prime &=&{\Scr A}\Omega=\{-\frac{1+st}{e_0^2},\, s,\, t,\,
\frac{(1-st)(e_0^1-e_0^2 u)}{e_0^2 e_1^1},\, st (-e_0^1+e_0^2 u
),\, tu,\, su,\, -e_1^1 st\}\,.\nonumber\\&&
\end{eqnarray}
Conditions (\ref{minimum}) and (\ref{susyminima}) are clearly
satisfied by the same values of the moduli (\ref{vac}). On this
vacuum in the new basis $X^{\prime 0}=0$. It seems that if, in
Michelson's basis, we have both electric and  magnetic charges
 in the direction of $X^{\prime 0}\neq 0$ (graviphoton)
supersymmetry is broken.  Therefore in the symplectic basis
$\Omega^\prime$ we can use special coordinates to describe the
supersymmetric vacuum \eq{vac}, in a patch $X^{\prime i}\neq 0$
only if $i\neq 0$. In what follows we shall consider the patch
$X^{\prime 1}\neq 0$. We refer the reader to Appendix B, eq.
\eq{nnew}, for the explicit form of the period matrix in
Michelson's basis at $s=-1/t$ and $\tau=-u$. Note that the
expression of the components $\mathcal{N}_{0,0}$ and
$\mathcal{N}_{0,3}$ in eq. \eq{nnew} are consistent with
conditions (\ref{ncond}), recalling that in this case
$m^{10}=2/e_0^2$.\par Let us write the prepotential in Michelson's
basis as a function of $s,t,u$:
\begin{eqnarray}
{\Scr F}=\frac{1}{2}\,X^{\prime
\Lambda}F^\prime_{\Lambda}&=&\frac{t}{s^2}\,\left(
\frac{{e_0^1}\,t}{{e_0^2}} + \left( \frac{1}{s} - t \right) \,u
\right) \,.
\end{eqnarray}
We may express the above prepotential in terms of new special
coordinates $s^\prime,t^\prime,u^\prime$ in the patch $X^{\prime 1
}=1$:
\begin{eqnarray}
s^\prime &=& \frac{X^{\prime 2}}{X^{\prime 1}}\,\,;\,\,\,t^\prime
= \frac{X^{\prime 0}}{X^{\prime 1}}\,\,;\,\,\,u^\prime=
\frac{X^{\prime 3}}{X^{\prime 1}}\,.
\end{eqnarray}
We refer the reader to eq. \eq{stuprime} of Appendix B for the
explicit form of these coordinates as functions of the old ones
$s,t,u$. The prepotential in these variables is:
\begin{eqnarray}
{\Scr F}&=&\frac{{e_0^1}\,{s^\prime}}{{e_0^2}} +
\frac{{e_0^2}\,{e_1^1}\,{t^\prime}\,{u^\prime}}{2} -
\frac{{e_1^1}\,{\sqrt{-4\,{s^\prime} +
{({e_0^2})}^2\,{{t^\prime}}^2}}\,{u^\prime}}{2}
\end{eqnarray}
One may check that:
\begin{eqnarray}
F_0^\prime &=&\partial_{t^\prime}{\Scr F}\,\,;\,\,\,F_2^\prime =\partial_{s^\prime}{\Scr
F}\,\,;\,\,\,F_3^\prime
=\partial_{u^\prime}{\Scr F}\,,\nonumber\\
F_1^\prime &=& 2\,{\Scr F}-t^\prime\,\partial_{t^\prime}{\Scr
F}-s^\prime\,\partial_{s^\prime}{\Scr F}-u^\prime\,\partial_{u^\prime}{\Scr F}\,.
\end{eqnarray}
\section{Cubic prepotentials}
Let us consider a special K\"ahler geometry with a generic cubic
prepotential:
\begin{eqnarray}
{\Scr F}&=&\frac{1}{6}\, \kappa_{ijk}\,z^i\,z^j\,z^k\,.
\end{eqnarray}
Let us denote the real components of $z^i$ as $z^i=x^i+i \,
\lambda^i$. The metric has the form:
\begin{eqnarray}
G_{ij}&=&-\frac{3}{2}\,\left(\frac{\kappa_{ij}}{\kappa}-\frac{3}{2}\,\frac{\kappa_i\,\kappa_j}{\kappa^2}\right)\,,
\label{gab}
\end{eqnarray}
where
\begin{eqnarray}
\kappa&=&\kappa_{ijk}\,\lambda^i\,
\lambda^j\,\lambda^k\,\,\,;\,\,\,\, \kappa_i=\kappa_{ijk}\,
\lambda^j\,\lambda^k\,\,\,;\,\,\,\,\kappa_{ij}=\kappa_{ijk}\,\lambda^k\,.
\end{eqnarray}
The real and imaginary components of the period matrix
$\mathcal{N}_{\Lambda\Sigma}$ are then computed to be:
\begin{eqnarray}
{\rm
Re}(\mathcal{N})&=&\left(\matrix{\frac{1}{3}\,\kappa_{ijk}\,x^i\,
x^j\, x^k& -\frac{1}{2}\,\kappa_{ijk}\, x^j\,
x^k\cr-\frac{1}{2}\,\kappa_{ijk}\,x^j\,x^k & \kappa_{ijk}\,x^k
}\right)\,,\nonumber\\
{\rm
Im}(\mathcal{N})&=&\frac{1}{6}\,\kappa\,\left(\matrix{1+4\,G_{ij}\,
x^i\,x^j & -4\,G_{ij}\, x^j\cr -4\,G_{ij}\, x^j & 4\,G_{ij}
}\right)\,.\label{Ncubic}
\end{eqnarray}
The positivity domain of the Lagrangian requires $\kappa<0$.\par
As we have seen for Michelson's basis of charges where
$e_{0},m^0\neq 0$, the existence of supersymmetric vacua implies
$X^0=0$. Consider the case of a cubic prepotential and charges in
Michelson's basis, but with no charge along the $0$--direction:
\begin{eqnarray}
e_{i_0}&=&e_{i_0}^2\,\tau+e_{i_0}^1\,\,\,;\,\,\,\,\,e_{j}=e_j^1\,;\,\,i_0,\,j\neq 0\,\,;\,i_0\neq j\,,\nonumber\\
m^{i_0}&=& m^{1\,i_0}\,.
\end{eqnarray}
The minimum conditions \eq{ncond} becomes:
\begin{eqnarray}
\overline{\mathcal{N}}_{i_0i_0}&=&\frac{e_{i_0}^1+\tau\,e_{i_0}^2}{m^{1i_0}}\,\,;\,\,\,\overline{\mathcal{N}}_{i_0j}=
\frac{e_j^1}{m^{1i_0}}\,\,;\,\,\,\overline{\mathcal{N}}_{i_0k}=0\,\,\,k\neq
i_0,\,j \,.\label{ncond2}
\end{eqnarray}
 Using eqs.
(\ref{Ncubic}) we can write the minimum conditions \eq{ncond2} as
follows:
\begin{eqnarray}
\mbox{conditions on ${\rm
Im}(\mathcal{N})$}&:&\cases{G_{i_0i_0}=-\frac{3}{2\,\kappa}\,\frac{e_{i_0}^2}{m^{1\,i_0}}\,\tau_2\,,\cr
G_{i_0k}=0\,\,\,k\neq i_0\,,\cr G_{i_0 k}\, x^k=0\,,}\label{condI}\\
\mbox{conditions on ${\rm
Re}(\mathcal{N})$}&:&\cases{\kappa_{i_0i_0 k }\, x^k
=\frac{e_{i_0}^2\,\tau_1+e_{i_0}^1}{m^{1\,i_0}}\,,\cr \kappa_{i_0j
k }\, x^k =\frac{e_{j}^1}{m^{1\,i_0}}\,,\cr \kappa_{i_0k l  }\,
x^l=0\,\,\,k\neq i_0,\,j\,,\cr \kappa_{i_0 kl}\, x^k\,
x^l=0\,.}\label{condR}
\end{eqnarray}
From eqs. \eq{condI},  \eq{condR} it follows that, if
$e_{i_0},\,e_j\neq 0$:
\begin{eqnarray}
x^{i_0}&=&x_j=0 \,.
\end{eqnarray}
If we further require supersymmetry we need to impose:
\begin{eqnarray}
X^{i_0}&=&0\,.\label{susy}
\end{eqnarray}
In the special coordinate basis there are components $X^{i_0}$
which can vanish, their imaginary part should not correspond to
Cartan isometries, e.g. brane coordinates.\par Let us specialize
to cubic prepotentials defining homogeneous spaces. The general
form of ${\Scr F}$ is given in \cite{dwvp}:
\begin{eqnarray}
{\Scr
F}&=&\frac{1}{2}\,[z^1\,(z^2)^2-z^1\,(z^\mu)^2-z^2\,(z^u)^2+\gamma_{\mu
uv}\,z^\mu\,z^u\,z^v ]\,,\label{fgen}
\end{eqnarray}
where $z^\mu=\{z^3,z^\alpha\}$ is a vector in the fundamental of
${\rm SO}(1+q)$, $z^u=\{z^r,\,z^n\}$ ($u=1,\dots, 2\,d_s$)
transforms in the spinorial representation of ${\rm SO}(1+q)$,
$z^r,\,z^n$ being the chiral components with respect to ${\rm
SO}(q)$, and $\gamma_\mu$ are the generators of the corresponding
Clifford algebra. The expression \eq{fgen} can be recast in the
following form:
\begin{eqnarray}
{\Scr F}&=&stu-\frac{s}{2}\,(z^n)^2
-\frac{u}{2}\,(z^r)^2-\frac{t}{2}\,(z^\alpha)^2+\gamma_{\alpha
k r}\,z^\alpha\, z^n\,z^r\,,\nonumber\\
\alpha &=&1,\dots,q\,\,;\,\,\,n=1,\dots,
d_s\,\,;\,\,\,r=1,\dots,d_s\,,
\end{eqnarray}
if we identify $s=z^2+z^3$, $u=z^2-z^3$ and $t=2\,z^1$.\par Since $s,t,u$ are moduli whose
imaginary parts are related to Cartan isometries, namely ${\rm Im}(s)=-e^{\phi_s},\,{\rm
Im}(t)=-e^{\phi_t},\,{\rm Im}(u)=-e^{\phi_u}$, they cannot be set to zero. Only the remaining
moduli $z^\alpha,\,z^n,\,z^r$ can be set to zero, and thus, in the study of supersymmetric
vacua, we shall consider three different cases in which the charges $e_{i_0},\,m^{i_0}$ are
chosen along the directions $X^\alpha=z^\alpha,\,X^n=z^n,\,X^r=z^r$.
\paragraph{Case $i_0=\bar{\alpha}$.} Let us start from the
conditions (\ref{condR}). The first equation gives
\begin{eqnarray}
\kappa_{\bar{\alpha}\bar{\alpha}t}\,x^t=\frac{1}{m^{2\,\bar{\alpha}}}(e_{\bar{\alpha}}^1\,\tau_1+e_{\bar{\alpha}}^2)\,.\label{rt}
\end{eqnarray}
The remaining conditions depend of the choice of the index $j$ of the additional electric
charge $e_j$. Choosing $j=\beta\neq \bar{\alpha}$ or $j=s,t,u$, conditions (\ref{condR}) imply
$e_j=0$. The only cases in which $e_j$ can be non--vanishing correspond to:
\begin{eqnarray}
j&=&\bar{n}\,\Rightarrow \, \cases{\kappa_{\bar{\alpha}\,n\,r}\,
x^r=\frac{e_n^1}{m^{1\,\bar{\alpha}}}\,\delta_{n\bar{n}}\cr \kappa_{\bar{\alpha}\,r\,n}\,
x^n=0}\,,\nonumber\\
j&=&\bar{r}\,\Rightarrow \, \cases{\kappa_{\bar{\alpha}\,r\,n}\,
x^n=\frac{e_j^1}{m^{1\,\bar{\alpha}}}\,\delta_{r\bar{r}}\cr \kappa_{\bar{\alpha}\,n\,r}\,
x^r=0}\,.
\end{eqnarray}
The last of conditions (\ref{condR}) does not imply any new constraint.\par Let us now
consider the implications of conditions (\ref{condI}). From eq. (\ref{gab}) we can write:
\begin{eqnarray}
G_{\bar{\alpha}k}&=&-\frac{3}{2}\,\left(\frac{\kappa_{\bar{\alpha}k}}{\kappa}
-\frac{3}{2}\,\frac{\kappa_{\bar{\alpha}}\,\kappa_k}{\kappa^2}\right)\,,
\end{eqnarray}
we may distinguish two cases: $\kappa_{\bar{\alpha}}=0$ and
$\kappa_{\bar{\alpha}}\neq 0$. In the former case vanishing of
$G_{\bar{\alpha}t}=0$, which is satisfied if
$\kappa_{\bar{\alpha}t}=\kappa_{\bar{\alpha}\bar{\alpha}t}\,\lambda^{\bar{\alpha}}=0$
which in turn implies $\lambda^{\bar{\alpha}}=0$. This latter
condition, together with $x^{\bar{\alpha}}=0$ from eqs.
(\ref{condI}), fixes $X^{\bar{\alpha}}=0$ and thus \emph{the
vacuum is supersymmetric}. The remaining conditions in eqs.
(\ref{condI}) imply:
\begin{eqnarray}
\kappa_{\bar{\alpha}\bar{\alpha}t}\,\lambda^t
&=&\frac{e_{\bar{\alpha}}^2}{m^{1\,\bar{\alpha}}}\,\tau_2\,,\label{it}\\
0&=&\kappa_{\bar{\alpha} n}=\kappa_{\bar{\alpha} n r}\, \lambda^r\,,\\
0&=&\kappa_{\bar{\alpha} r}=\kappa_{\bar{\alpha} n r}\,
\lambda^n\,.
\end{eqnarray}
Eqs. (\ref{rt}), (\ref{it}) imply that the complex scalar $t$ is fixed to the complex value:
\begin{eqnarray}
t&=&t_0=\frac{e_{\bar{\alpha}}}{\kappa_{\bar{\alpha}\bar{\alpha} t}\, m^{\bar{\alpha}}}\,.
\end{eqnarray}
The scalars $z^{\beta}$, $\beta\neq \bar{\alpha}$ are moduli in this supersymmetric
vacuum.\par Relaxing condition $\kappa_{\bar{\alpha}}=0$ which imply unbroken supersymmetry,
we obtain  involved non--linear equations to be solved. We shall not discuss here the most
general solution of these equations.
\paragraph{Case $i_0=\bar{n}$.}
The first of eqs. (\ref{condR}) gives
\begin{eqnarray}
\kappa_{\bar{n}\bar{n}s}\,x^s=\frac{1}{m^{2\,\bar{n}}}(e_{\bar{n}}^1\,\tau_1+e_{\bar{n}}^2)\,.\label{rs}
\end{eqnarray}
Let us discuss the remaining conditions for various choices of the
electric charge $e_j$. For $j=n\neq \bar{n}$ or $j=s,t,u$,
conditions (\ref{condR}) imply $e_j=0$. The only cases allowing
non--vanishing $e_j$ correspond to:
\begin{eqnarray}
j&=&\bar{\alpha}\,\Rightarrow \, \cases{\kappa_{\bar{n}\,\alpha\,r}\,
x^r=\frac{e_j^1}{m^{1\,\bar{n}}}\,\delta_{\alpha\bar{\alpha}}\cr \kappa_{\bar{n}\,r\,\beta}\,
x^\beta=0}\,,\nonumber\\
j&=&\bar{r}\,\Rightarrow \, \cases{\kappa_{\bar{n}\,r\,\beta}\,
x^\beta=\frac{e_j^1}{m^{1\,\bar{n}}}\,\delta_{r\bar{r}}\cr \kappa_{\bar{n}\,\beta\,s}\,
x^s=0}\,.
\end{eqnarray}
The last of conditions (\ref{condR}) does not imply any new constraint.\par As far as
conditions  (\ref{condI}) are concerned, the relevant components of the metric are:
\begin{eqnarray}
G_{\bar{n}i}&=&-\frac{3}{2}\,\left(\frac{\kappa_{\bar{n}i}}{\kappa}
-\frac{3}{2}\,\frac{\kappa_{\bar{n}}\,\kappa_i}{\kappa^2}\right)\,.
\end{eqnarray}
We start discussing the  $\kappa_{\bar{n}}=0$ case. The vanishing
of $G_{\bar{\alpha}s}=0$, which is satisfied if
$\kappa_{\bar{\alpha}s}=0$ implies $\lambda^{\bar{n}}=0$. This
condition, together with $x^{\bar{n}}=0$ from eqs. (\ref{condI}),
fixes $X^{\bar{n}}=0$ and thus ensures supersymmetry of the
vacuum. The remaining conditions in eqs. (\ref{condI}) imply:
\begin{eqnarray}
\kappa_{\bar{n}\bar{n} s}\,\lambda^s
&=&\frac{e_{\bar{n}}^2}{m^{1\,\bar{n}}}\,\tau_2\,,\label{is}\\
0&=&\kappa_{\bar{n} \alpha}=\kappa_{\bar{n} \alpha r}\, \lambda^r\,,\\
0&=&\kappa_{\bar{n}r}=\kappa_{\bar{n} r \alpha}\,
\lambda^\alpha\,.
\end{eqnarray}
Eqs. (\ref{rs}),(\ref{is}) imply that the complex scalar $s$ is fixed to the complex value:
\begin{eqnarray}
s&=&s_0=\frac{e_{\bar{n}}}{\kappa_{\bar{n}\bar{n} s}\, m^{\bar{n}}}\,.
\end{eqnarray}
The scalars $z^{n}$, $n\neq \bar{n}$ are moduli in this supersymmetric vacuum.\par Also in
this case relaxing condition $\kappa_{\bar{n}}=0$, which imply unbroken supersymmetry, we have
to solve involved non--linear conditions. We shall not discuss here the existence of a
non--trivial solution.\par
\paragraph{Case $i_0=\bar{r}$.} This case is analogous to the previous
one upon substituting $r\leftrightarrow n$ and $s\leftrightarrow
u$.\par We now show that in the spacial cases
of$L(0,P,\dot{P}),\,\,L(q,0)$ manifolds, all vacua are
supersymmetric. Consider first the $q=0$ case defining the
$L(0,P,\dot{P})$ manifold. If we take $i_0=\bar{k}$ from eqs.
(\ref{condI}) we derive
\begin{eqnarray}
G_{\bar{k}t}&=&\frac{9}{2}\,\frac{\kappa_{\bar{k}}\,\lambda^s\,\lambda^u}{\kappa^2}=0\rightarrow
\kappa_{\bar{k}}=0\,\Rightarrow\,\lambda^{\bar{k}}=0\,,
\end{eqnarray}
the last condition, together with $x^{\bar{k}}=0$ ensures supersymmetry of the vacuum.\par
Similarly if we take $i_0=\bar{r}$, from $G_{\bar{r}t}=0$ we derive $\lambda^{\bar{r}}=0$ and
thus that the vacuum is supersymmetric.\par The same arguments apply to the $L(q,0)$.
\paragraph{${\rm Sp}(6)/{\rm U}(3)$ example.} The coordinates of
the six--dimensional special K\"ahler manifold are given by the
independent entries of a symmetric ${\rm U}(3)$ complex tensor
$Z^{ij},\,i,j=1,2,3$. Choosing:
\begin{eqnarray}
z^1&=&Z^{11}\,\,;\,\,\,z^2=Z^{22}\,\,;\,\,\,z^3=Z^{33}\,\,;\,\,\,z^4=-Z^{23}\,\,;\,\,\,z^5=-Z^{13}\,\,;\,\,\,z^6=-Z^{12}\,,\nonumber\\&&
\end{eqnarray}
the prepotential ${\Scr F}$ has the form of the following ${\rm
U}(3)$--invariant polynomial
\begin{eqnarray}
{\Scr F}&=&\frac{1}{6}\,\epsilon_{ijk}\epsilon_{lmn}\, Z^{il}\,
Z^{jm}\,
Z^{kn}=z^1\,z^2\,z^3-z^1\,(z^4)^2-z^2\,(z^5)^2-z^3\,(z^6)^2-2\,z^4\,z^5\,z^6\,.\nonumber\\&&\label{ff}
\end{eqnarray}
Equation (\ref{ff}) is consistent with the general form of the
cubic polynomial for homogeneous manifolds given in \cite{dwvp1},
which, for the present $L(1,1)$ case, reads:
\begin{eqnarray}
\kappa(h)&=&6\,[h^1\,(h^2+h^3)\,(h^2-h^3)-(h^2-h^3)\,(h^5)^2-(h^2+h^3)\,(h^6)^2-h^1\,(h^4)^2-\nonumber\\&&-
2\,h^4\,h^5\,h^6]\,.\nonumber\\&&
\end{eqnarray}
In this case we may identify the coordinates $s,t,u$ parametrizing
the $[{\rm SU}(1,1)/{\rm U}(1)]^3$ submanifold, with $z^1, z^2,
z^3$ respectively, and $z^\alpha=z^5,\,z^k=z^4,\,z^r=z^6$.
Consider taking $e_{i_0},\,m^{i_0}$ along the direction $k=4$.
Conditions (\ref{condI}) imply:
\begin{eqnarray}
G_{4s}&=&-\frac{18}{\kappa^2}\,(\lambda_4\,\lambda_5+\lambda_u\,\lambda_6)\,(\lambda_4\,\lambda_6+\lambda_t\,\lambda_5)=0\,,\nonumber\\
G_{4t}&=&-\frac{18}{\kappa^2}\,(\lambda_s\,\lambda_u+\lambda_5^2)\,(\lambda_4\,\lambda_s+\lambda_6\,\lambda_5)=0\,,\nonumber\\
G_{4u}&=&-\frac{18}{\kappa^2}\,(\lambda_s\,\lambda_t+\lambda_6^2)\,(\lambda_4\,\lambda_s+\lambda_6\,\lambda_5)=0\,,\nonumber\\
G_{45}&=&\frac{18}{\kappa^2}\,(2\,{{\lambda }_1}\,{{\lambda
}_2}\,{{\lambda }_4}\,{{\lambda }_5} +
  {{\lambda }_1}\,{{\lambda }_2}\,{{\lambda }_3}\,{{\lambda }_6} +
  {{\lambda }_1}\,{{{\lambda }_4}}^2\,{{\lambda }_6} + {{\lambda }_2}\,{{{\lambda }_5}}^2\,{{\lambda }_6} -
  {{\lambda }_3}\,{{{\lambda }_6}}^3)=0\,,\nonumber\\
  G_{46}&=&\frac{18}{\kappa^2}\,({{\lambda }_1}\,{{\lambda }_2}\,{{\lambda }_3}\,{{\lambda }_5} +
  {{\lambda }_1}\,{{{\lambda }_4}}^2\,{{\lambda }_5} - {{\lambda }_2}\,{{{\lambda }_5}}^3 +
  2\,{{\lambda }_1}\,{{\lambda }_3}\,{{\lambda }_4}\,{{\lambda }_6} +
  {{\lambda }_3}\,{{\lambda }_5}\,{{{\lambda }_6}}^2)=0\,,\nonumber\\
    G_{44}&=&\frac{18}{\kappa^2}\,({{{\lambda }_1}}^2\,{{\lambda }_2}\,{{\lambda }_3} + {{{\lambda }_1}}^2\,{{{\lambda }_4}}^2 -
  {{\lambda }_1}\,{{\lambda }_2}\,{{{\lambda }_5}}^2 +
  2\,{{\lambda }_1}\,{{\lambda }_4}\,{{\lambda }_5}\,{{\lambda }_6} -
  {{\lambda }_1}\,{{\lambda }_3}\,{{{\lambda }_6}}^2 + 2\,{{{\lambda }_5}}^2\,{{{\lambda
  }_6}}^2)=\nonumber\\&=&-\frac{3}{2\,\kappa}\,\frac{e^1_{i_0}}{m^{2i_0}}\,\tau_2\,.
\end{eqnarray}
According to our general analysis condition
\begin{eqnarray}
\kappa_4&=&-4\,(\lambda_2\,\lambda_4+\lambda_5\,\lambda_6)=0\,,
\end{eqnarray}
characterizes the supersymmetric vacuum which always exists. In this case there are no other
solutions.
\section{Acknowledgments}
 R.D., M.T. and S.V.  would like to thank the Physics
Department of CERN, where part of this work was done, for its kind hospitality. S.F. would
like to thank the Physics Department of Politecnico di Torino for its kind hospitality and the
excellent cooking of the restaurant ``Serendip''.
\par
 Work
supported in part by the European Community's Human Potential Program under contract
MRTN-CT-2004-005104 `Constituents, fundamental forces and symmetries of the universe', in
which R. D'A. and M.T.  are associated to Torino University. The work of S.F. has been
supported in part by European Community's Human Potential Program under contract
MRTN-CT-2004-005104 `Constituents, fundamental forces and symmetries of the universe', in
association with INFN Frascati National Laboratories and by D.O.E. grant DE-FG03-91ER40662,
Task C. The work of S.V. has been supported by DFG -- The German Science Foundation, DAAD --
the German Academic Exchange Service and by the European Community's Human Potential Program
under contract  HPRN-CT-2000-00131.
\appendix{}
\section{Appendix}

In the present appendix we recall our notations. The metric of the
original quaternionic manifold is \cite{fs}:
\begin{eqnarray}
ds^2&=&h_{\hat{u}\hat{v}}\,dq^{\hat{v}}=K_{Q\,a\bar{b}}\,
\partial_\mu w^a\partial^\mu
\bar{w}^{\bar{b}}+(\partial\varphi)^2+ \frac{e^{4\,\varphi}}{4}\,(\partial a -V\times
\partial V)^2-\nonumber\\&&-\frac{e^{2\,\varphi}}{2}\,\partial_\mu V\, {\Scr
M}\,\partial^\mu V\,,\label{qmet}
\end{eqnarray}
where $K_{Q\,a\bar{b}}=\partial_a\partial_{\bar{b}}\,K_Q$ and  we have denoted the scalar
fields by $\{\varphi,\,a,\,w^a,\,\xi^\Lambda,\,\tilde{\xi}_\Lambda\}$\footnote{In the present
paper we have chosen to denote the axions deriving from the RR forms by the letter $\xi$
instead of $\zeta$, which is more often used in the literature, in order not to create
confusion with the hyperinos.}, $V$ in \eq{qmet} is the symplectic vector defined as:
\begin{eqnarray}
V&=&\{\xi^\Lambda,\,\tilde{\xi}_\Lambda\}\,,
\end{eqnarray}
and $``\times"$ denotes the symplectic invariant scalar product:
\begin{eqnarray}
V\times W&=&V^\L W_\L-V_\L W^\L\,.
\end{eqnarray}
The matrix ${\Scr M}$ in \eq{qmet} is negative definite and has the following form:
\begin{eqnarray}
{\Scr M}&=&\left(\matrix{\mathcal{R}\,\mathcal{I}^{-1}\,\mathcal{R}+\mathcal{I}&
\mathcal{R}\,\mathcal{I}^{-1}\cr \mathcal{I}^{-1}\,\mathcal{R}& \mathcal{I}^{-1} }\right)\,,
\end{eqnarray}
 where $\mathcal{R}$ and $\mathcal{I}$ are the real and
imaginary parts $\mathcal{R}={\rm Re}(\mathcal{M})$, $\mathcal{I}={\rm Im}(\mathcal{M})$ of
the ``period matrix" $\mathcal{M}_{\Lambda\Sigma}$ associated to the special K\"ahler
submanifold parametrized by $z^a$.\par The scalar fields dual to the NSNS and RR tensors
$B_{\mu\nu},\,C_{\mu\nu}$ are $a,\,\tilde{\xi}_0$ respectively while $\xi^a$ and
$\tilde{\xi}_a$, $a=1,\dots\ h_{1,1}$, are the remaining RR scalars originating from the
2--form and the 4--form respectively. Finally $\xi^0$ corresponds to the ten dimensional
axion, $\varphi$ is the four dimensional dilaton and $w^a$ are the K\"ahler moduli. The metric
$M_{IJ}=h_{IJ}$, where the values $I,J=1,2$ label the scalars $a,\,\tilde{\xi}_0$
respectively, and its inverse $M^{IJ}$ have the following form:
\begin{eqnarray}
M_{IJ}&=&\frac{e^{4\,\varphi}}{4}\,\left(\matrix{1 & -\xi^0\cr
-\xi_0 & (\xi^0)^2-2\, e^{-2\,\varphi}
\mathcal{I}^{-1|\,00}}\right)\,,\nonumber\\
M^{IJ}&=&-\frac{2}{\mathcal{I}^{-1|\,00}}\,e^{-2\,\varphi}\,\left(\matrix{(\xi^0)^2-2\,
e^{-2\,\varphi}\, \mathcal{I}^{-1|\,00} & \xi^0\cr \xi_0 & 1}\right)\,,\label{MM}
\end{eqnarray}
 Note that the expression of $M_{IJ}$ coincides
with that of $\omega^x_I\omega^x_J$ given by:
\begin{eqnarray}
\omega^x_I\omega^x_J&=&\frac{e^{4\,\varphi}}{4}\,\left(\matrix{1 &
-\xi^0\cr -\xi_0 & (\xi^0)^2+16\, e^{-2\,\varphi+K_Q} }\right)\,,
\end{eqnarray}
\emph{only} in the case of  cubic quaternionic geometries for which the following relation
holds: \begin{eqnarray} e^{K_Q}&=&-\frac{1}{8}\,\mathcal{I}^{-1|\,00}\,.
\end{eqnarray}
 Using the explicit metric \eq{qmet} and
eqs. \eq{MM} we can now compute the quantities $A_u^I$:
\begin{eqnarray}
A_u^I
dq^u=M^{IJ}\,h_{Iu}\,dq^u&=&\frac{1}{\mathcal{I}^{-1|00}}\,\left[\left(\matrix{-\mathcal{I}^{-1|00}\,\xi^a+\xi^0\,
\mathcal{I}^{-1|0a} \cr \mathcal{I}^{-1|0a}
}\right)\,d\tilde{\xi}_a+\right.\nonumber\\&&\left.+\left(\matrix{\mathcal{I}^{-1|00}\,\tilde{\xi}_\Lambda+\xi^0\,
(\mathcal{R}\mathcal{I}^{-1})_\Lambda{}^0\cr (\mathcal{R}\mathcal{I}^{-1})_\Lambda{}^0
}\right)\,d\xi^\Lambda\right]\,.
\end{eqnarray}
If we redefine $a\rightarrow a-\xi^\Lambda\,\tilde{\xi}_\Lambda$
the metric will no more depend on $\tilde{\xi}$ and $A^I_u$ will
have the form:
\begin{eqnarray}
A_u^I
dq^u&=&\frac{1}{\mathcal{I}^{-1|00}}\,\left[\left(\matrix{-2\,\mathcal{I}^{-1|00}\,\xi^a+2\,\xi^0\,\mathcal{I}^{-1|0a}
\cr \mathcal{I}^{-1|0a} }\right)\,d\tilde{\xi}_a+\left(\matrix{2\,\xi^0\,
(\mathcal{R}\mathcal{I}^{-1})_\Lambda{}^0\cr (\mathcal{R}\mathcal{I}^{-1})_\Lambda{}^0
}\right)\,d\xi^\Lambda\right]\,.
\end{eqnarray}
 Let us define the following forms:
\begin{eqnarray}
v&=&\frac{1}{2}\,e^{2\,\varphi}\,[-2\,e^{-2\,\varphi}\,d\varphi-i\,(da+\tilde{\xi}^T\,
d\xi-\xi^T\, d\tilde{\xi})]\,,\nonumber\\
u&=&i\,e^{\varphi+\frac{K_Q}{2}}\, Z^T\,(\overline{{\mathcal
M}}\,d\xi+d\tilde{\xi})\,, \nonumber\\
E&=&i\,e^{\varphi-\frac{K_Q}{2}}\,P\,N^{-1}\,(\overline{{\mathcal
M}}\,d\xi+d\tilde{\xi})\,, \nonumber\\
e&=&P\,dZ\,,\label{uvforms}
\end{eqnarray}
where $Z^\Lambda=\{1,w^a\}$ and the matrices $P$ and $N$ are defined as follows:
\begin{eqnarray}
P^{\underline{a}}{}_0 &=&-e_b{}^{\underline{a}}{} Z^b\,\,\,;\,\,\,\,P^{\underline{a}}{}_b=e_b{}^{\underline{a}}\,\,\,
(b,\,a=1,\dots, h_{2,1})\,,\\
N_{\Lambda\Sigma}&=&\frac{1}{2}{\rm Re}(\frac{\partial^2 {\Scr
F}_Q}{\partial Z^\Lambda\partial Z^\Sigma})\,.
\end{eqnarray}
${\Scr F}_Q$ being the prepotential of the special K\"ahler
manifold  embedded in the quaternionic manifold,
$e_a{}^{\underline{b}}$ being the corresponding vielbein (the
underlined indices are the rigid ones). One can check that in
terms of the forms in \eq{uvforms}, the metric \eq{qmet} has the
simple expression:
\begin{eqnarray}
ds^2&=&v\otimes \overline{v}+u\otimes \overline{u}+E\otimes \overline{E}+e\otimes
\overline{e}\,.
\end{eqnarray}
Let us now give the expression for the vielbein $\mathcal{U}$. In the heterotic case we have:
\begin{eqnarray}
\mathcal{U}^{1\dot \alpha}&=&\left(\matrix{v\cr
e^{\underline{a}}}\right)\,\,;\,\,\,\mathcal{U}^{1\dot \alpha}=\left(\matrix{u\cr
E^{\underline{a}}}\right)\,.
\end{eqnarray}
For the $O5/O9$ case we simply exchange $\mathcal{U}^{1\dot \alpha}\leftrightarrow
\mathcal{U}^{2\dot \alpha}$. In particular we can compute the components of
$\mathcal{U}^{A\dot\alpha}_I$ where $I=1$ is the component along $da$ and $I=2$ along
$d\tilde{xi}_0$, $a,\,\tilde{\xi}_0$ being the scalars dual to $B_{1\mu\nu}$ and $B_{2\mu\nu}$
respectively. We obtain in the heterotic case:
\begin{eqnarray}
\mathcal{U}^{1\dot \alpha}_{I=1}=-\frac{i}{2}\,e^{2\,\varphi}\,\left(\matrix{1\cr {\bf
0}_{(n_H-1)}}\right)\,\,&;&\,\,\,\mathcal{U}^{1\dot
\alpha}_{I=2}=\frac{i}{2}\,e^{2\,\varphi}\,\left(\matrix{\xi^0\cr {\bf
0}_{(n_H-1)}}\right)\,,\nonumber\\
\mathcal{U}^{2\dot \alpha}_{I=1}=\left(\matrix{0\cr {\bf
0}_{(n_H-1)}}\right)\,\,\,\,\,\,\,\,\,\,\,\,\,\,\,\,\,\,\,\,\,\,\,&;&\,\,\,\mathcal{U}^{2\dot
\alpha}_{I=2}=i\,e^{\varphi+\frac{K_Q}{2}}\,\left(\matrix{1\cr
e^{-{K_Q}}\,P^{\underline{a}}{}_\Lambda\,N^{-1|\Lambda 0}}\right)\,,
\end{eqnarray}
where the first entry of the above vectors corresponds to $\dot \alpha=0$. We note that in the
heterotic case $\xi^0=0$ so that $\mathcal{U}^{1\dot \alpha}_{I=2}=\mathcal{U}^{2\dot
\alpha}_{I=1}=0$, consistently with equations (\ref{uzero}), (\ref{zeru2}). In the $O5/O9$
case, exchanging $\mathcal{U}^{1\dot \alpha}\leftrightarrow \mathcal{U}^{2\dot \alpha}$ we
obtain the corresponding conditions.
\section{Special K\"ahler geometry in two different symplectic bases.}
The matrix ${\Scr A}$ relating the two relevant symplectic bases
$Q_M$ and $Q_L$ is:
\begin{eqnarray}
{\Scr A}&=&\left(\matrix{ - \frac{1}{{e_0^2}}  & 0 & 0 & 0 & 0 & 0
& 0 & - \frac{1}{{e_0^2}} \cr 0 & 1 & 0 & 0 & 0 & 0 & 0 & 0 \cr 0
& 0 & 1 & 0 & 0 & 0 & 0 & 0 \cr \frac{{e_0^1}} {{e_0^2}\,{e_1^1}}
& 0 & 0 & -\frac{1}{{e_1^1}} & - \frac{1}{{e_1^1}}  & 0 & 0 & -
\frac{{e_0^1}} {{e_0^2}\,{e_1^1}}  \cr 0 & 0 & 0 & 0 & -{e_0^2 } &
0 & 0 & -{e_0^1 } \cr 0 & 0 & 0 & 0 & 0 & 1 & 0 & 0 \cr 0 & 0 & 0
& 0 & 0 & 0 & 1 & 0 \cr 0 & 0 & 0 & 0 & 0 & 0 & 0 & -{e_1^1} \cr
}\right)\,.\label{A}
\end{eqnarray}
 The period matrix in the special coordinate
basis is
\begin{eqnarray}
\overline{\mathcal{N}}_{0,0}&=&-\frac{s^2{\left(
{\overline{t}}u - t\, {\overline{u}} \right) }^2 \!+ {
{\overline{s}}}^2{\left( tu -  {\overline{t}} {\overline{u}}
\right) }^2\! - 2s {\overline{s}}\left( t^2\,u
{\overline{u}} + { {\overline{t}}}^2\,u {\overline{u}} + t\,
{\overline{t}}\,\left( u^2 - 4\,u\, {\overline{u}} + {
{\overline{u}}}^2 \right) \right)  }{2\left( s -  {\overline{s}}
\right) \left( t -  {\overline{t}} \right)\left( u -
{\overline{u}} \right)
}\,,\nonumber\\
\overline{\mathcal{N}}_{0,1}&=&\frac{\overline{s} tu-\overline{t}
su-\overline{u} ts+\overline{s}
\overline{t}\overline{u}}{2\,(s-\overline{s})}\,,\nonumber\\
\overline{\mathcal{N}}_{0,2}&=&\frac{\overline{s} tu-\overline{t}
su+\overline{u} ts-\overline{s}
\overline{t}\overline{u}}{2\,(s-\overline{s})}\,,\nonumber\\
\overline{\mathcal{N}}_{0,3}&=&\frac{\overline{s} tu+\overline{t}
su-\overline{u} ts-\overline{s}
\overline{t}\overline{u}}{2\,(s-\overline{s})}\,,\nonumber\\
\overline{\mathcal{N}}_{1,1}&=&-\frac{(t-\overline{t})(u-\overline{u})}{2\,(s-\overline{s})}\,,\nonumber\\
\overline{\mathcal{N}}_{1,2}&=&\frac{u+\overline{u}}{2}\,,\nonumber\\
\overline{\mathcal{N}}_{1,3}&=&\frac{t+\overline{t}}{2}\,,\nonumber\\
\overline{\mathcal{N}}_{1,1}&=&-\frac{(s-\overline{s})(u-\overline{u})}{2\,(t-\overline{t})}\,,\nonumber\\
\overline{\mathcal{N}}_{2,3}&=&\frac{s+\overline{s}}{2}\,,\nonumber\\\overline{\mathcal{N}}_{3,3}&=&-\frac{(s-\overline{s})(t-\overline{t})}{2\,(u-\overline{u})}\,.\label{nstu}
\end{eqnarray}
The period matrix in the new basis $\Omega^\prime$ at $s=-1/t$ and
$\tau=-u$ reads:
\begin{eqnarray}
\overline{\mathcal{N}}_{0,0}&=&\frac{1}{2}\,e_0^2\,(e_0^1-u e_0^2)\,,\nonumber\\
\overline{\mathcal{N}}_{0,1}&=&0\,,\nonumber\\
\overline{\mathcal{N}}_{0,2}&=&0\,,\nonumber\\
\overline{\mathcal{N}}_{0,3}&=&\frac{1}{2}\,e_0^2\,e_1^1\,,\nonumber\\
\overline{\mathcal{N}}_{1,1}&=&\frac{2\,t^2\,{{\overline{t}}}^2\,\left(
u - {\overline{u}} \right) \, \left( {e_0^1} -
{e_0^2}\,{\overline{u}} \right) }{{e_0^1}\, {\left( t -
{\overline{t}} \right) }^2 - {e_0^2}\,\left( t^2\,u +
{{\overline{t}}}^2\,u +
2\,t\,{\overline{t}}\,\left( u - 2\,{\overline{u}} \right)  \right) }\,,\nonumber\\
\overline{\mathcal{N}}_{1,2}&=&\frac{{e_0^2}\,{\overline{u}}\,
\left( t^2\,u + {{\overline{t}}}^2\,u -
2\,t\,{\overline{t}}\,{\overline{u}} \right)  - {e_0^1}\,\left(
-2\,t\,{\overline{t}}\,u + t^2\,{\overline{u}} +
{{\overline{t}}}^2\,{\overline{u}} \right) }{- {e_0^1}\, {\left( t
- {\overline{t}} \right) }^2   + {e_0^2}\,\left( t^2\,u +
{{\overline{t}}}^2\,u +
2\,t\,{\overline{t}}\,\left( u - 2\,{\overline{u}} \right)  \right) }\,,\nonumber\\
\overline{\mathcal{N}}_{1,3}&=&-
\frac{{e_0^2}\,{e_1^1}\,t\,{\overline{t}}\, \left( t +
{\overline{t}} \right) \,\left( u - {\overline{u}} \right) } {-
{e_0^1}\,{\left( t - {\overline{t}} \right) }^2 + {e_0^2}\,\left(
t^2\,u + {{\overline{t}}}^2\,u +
2\,t\,{\overline{t}}\,\left( u - 2\,{\overline{u}} \right)  \right) }\,,\nonumber\\
\overline{\mathcal{N}}_{2,2}&=&\frac{2\,\left( u - {\overline{u}}
\right) \, \left( {e_0^1} - {e_0^2}\,{\overline{u}} \right)
}{{e_0^1}\, {\left( t - {\overline{t}} \right) }^2 -
{e_0^2}\,\left( t^2\,u + {{\overline{t}}}^2\,u +
2\,t\,{\overline{t}}\,\left( u - 2\,{\overline{u}} \right)  \right) }\,,\nonumber\\
\overline{\mathcal{N}}_{2,3}&=&\frac{{e_0^2}\,{e_1^1}\,\left( t +
{\overline{t}} \right) \, \left( u - {\overline{u}} \right)
}{-{e_0^1}\, {\left( t - {\overline{t}} \right) }^2 +
{e_0^2}\,\left( t^2\,u + {{\overline{t}}}^2\,u +
2\,t\,{\overline{t}}\,\left( u - 2\,{\overline{u}} \right) \right)
}\,,\nonumber\\\overline{\mathcal{N}}_{3,3}&=&\frac{-
{e_0^2}\,{({e_1^1})}^2\, {\left( t - {\overline{t}} \right) }^2
}{2\, \left( - {e_0^1}\,{\left( t - {\overline{t}} \right) }^2 +
{e_0^2}\,\left( t^2\,u + {{\overline{t}}}^2\,u +
2\,t\,{\overline{t}}\,\left( u - 2\,{\overline{u}} \right) \right)
\right) }\,.\label{nnew}
\end{eqnarray}
In this basis we can define special coordinates referred to the
patch in which $X^\prime 1\neq 0$ (we have rescaled
$\Omega^\prime$ by s):
\begin{eqnarray}
s^\prime &=& \frac{X^{\prime 2}}{X^{\prime 1}}\,\,;\,\,\,t^\prime
= \frac{X^{\prime 0}}{X^{\prime 1}}\,\,;\,\,\,u^\prime=
\frac{X^{\prime 3}}{X^{\prime 1}}\,\nonumber\\
s&=&\frac{-\left( {e_0^2}\,{t^\prime} \right) \pm
{\sqrt{-4\,{s^\prime} + {({e_0^2})}^2\,{{t^\prime}}^2}}}{2\,
{s^\prime}}\,\,;\,\,\,t=\frac{-\left( {e_0^2}\,{t^\prime} \right)
\pm {\sqrt{-4\,{s^\prime} +
{({e_0^2})}^2\,{{t^\prime}}^2}}}{2}\,;\nonumber\\u&=&\frac{{e_0^1}}{{e_0^2}}
\pm \frac{{e_1^1}\,{u^\prime}} {{\sqrt{-4\,{s^\prime} +
{({e_0^2})}^2\,{{t^\prime}}^2}}}\,,\label{stuprime}
\end{eqnarray}
we shall use the first solution (with the``$+$'' sign).

\end{document}